\documentclass[aip,
 amsmath,amssymb,
 reprint,%
]{revtex4-1}

\usepackage[utf8]{inputenc}
\usepackage[T1]{fontenc}
\usepackage{graphicx}
\usepackage{geometry}
\usepackage[normalem]{ulem}
\usepackage{color} 
\usepackage{wrapfig}
\usepackage{tabularx,ragged2e}
\usepackage{multirow}
\usepackage{array}
\usepackage{hyperref}
\usepackage[dvipsnames]{xcolor}
\def\DEL#1{}        % deletions

\begin{document}

\title{Scaling laws in axisymmetric magnetohydrodynamic duct flows}

\author{ A. Poy\'e$^{1,2}$, O. Agullo$^{2}$, N. Plihon$^{1}$, W.J.T. Bos$^{3}$, V. Desangles$^{1}$ and G. Bousselin$^{1}$}

\affiliation{Univ Lyon, Ens de Lyon, Univ Claude Bernard, CNRS, Laboratoire de Physique, F-69342 Lyon, France}
\affiliation{Aix-Marseille Universit\'e, CNRS, PIIM, UMR 7345, Marseille, France}
\affiliation{LMFA, CNRS, Ecole Centrale de Lyon, Universit\'e de Lyon,  Ecully, France}

\begin{abstract}
We report on a numerical study of axisymmetric flow of liquid metal in a circular duct with rectangular cross-section. The flow is forced through the combination of an axial magnetic field and a radial current. Sweeping a wide range of forcing parameters, we identify the different regimes which characterize the flows and explicit the associate scaling laws. Experimental results are interpreted in the light of our numerical simulations.
\end{abstract}

\maketitle

%%%%%%%%%%%%%%%%%%%%%%%%%%%%%%%%%%%%%%%%%%%%%
%%%%%%%%%%%%%%%%%%%%%%%%%%%%%%%%%%%%%%%%%%%%%
\section{Introduction}
%%%%%%%%%%%%%%%%%%%%%%%%%%%%%%%%%%%%%%%%%%%%%
%%%%%%%%%%%%%%%%%%%%%%%%%%%%%%%%%%%%%%%%%%%%%

The magnetohydrodynamic (MHD) flow through a toroidal duct is one of the elementary academic flows to study the behavior of liquid metals. It proves experimentally convenient, since it is limited in size and of relatively low complexity, numerically convenient since no inlet-outlet conditions have to be specified, and analytically convenient due to its symmetry. The flow is electrically driven and has been proposed for magneto-rotational instability (MRI) experiments \cite{Khalzov2010, Baylis1971a}, MRI being an essential ingredient in the dynamics of accretion disks. Duct MHD flow gives however rise to a number of competing mechanisms due to the simultaneous presence of rotation and a magnetic field and its description, even in the laminar case, is far more involved than simple duct flow of a non-conducting fluid. A number of different instabilities can be triggered depending on the control-parameters and the mean flows are not even well identified.

The possibility to observe MRI instability in electrically driven annular duct flow and Taylor-Couette flow was investigated both experimentally and numerically\cite{Sisan2004, Stefani2006, Khalzov2010}. Further investigations of electrically driven duct flow have been reported in literature \cite{Baylis1964, Tabeling1982, Moresco2003, Moresco2004,Stelzer2015a, Stelzer2015b, Boisson2012}. One important observation in these works is that a large number of regimes can be observed, depending on the strength of the magnetic field, electric current, fluid properties and geometry of the duct.

Early studies \cite{Baylis1971a} focused on the diffusive regime, which is most easily treated analytically. The more complex case where inertial effects become important was treated more recently \cite{Khalzov2006,Khalzov2010}, both numerically and analytically. In a recent attempt \cite{Boisson2017} to sort out the different possible flows, experiments were carried out to distinguish between inertial and diffusive regimes, focusing on the influence of the interaction parameter and flow geometry.  The flows in this experiment, and in most other experimental devices with channel widths of the order of 10cm, are restricted to magnetic fields from 0.1 to 1~T and can reach a $H\!a \in 400, 4000$ \cite{Baylis1971a,Stelzer2015b}. Here, the Hartmann-number $H\!a$ measures the magnetic field strength compared to visco-resistive effects. 

Attaining an inertia dominated regime through magnetic forcing for large values of the Hartmann number requires a strong magnetic field (typically of the order from 1 to 10 Tesla) over centimetric distances, which remains a technical challenge, not attained in most studies, an exception being the reference work of Moresco and Alboussière \cite{Moresco2004}. The limit of low Hartmann number $H\!a \ll 1$ has been investigated only with electrolytes and small current \cite{Digilov2007,Suslov2017}. Some other non-dimensional numbers or criteria have been introduced to characterize the flow, such as the Prandtl number, the inertial number and the Reynolds number. It will become particularly clear that the observed mean flow does not follow always the underlying criteria associated with those non-dimensional numbers. %We aim to identify and caracterize the possible flow regimes. 

In this study, we aim to identify and characterize the possible flow regimes of electrically driven duct flows, for given fluid properties. For this purpose, we will carry out simulations over a wide range of Hartmann  (high/low magnetic field) and inertial number (high/low electrical drive), focusing on the specific case of axisymmetric flow patterns. 
More specifically, we characterize the various mean flows observed in this configuration and compare with existing or derived scaling laws. We present also a study of the influence of the geometry of the duct (aspect ratio of the cross section and mean radius of the duct) on the nature of the flows. Finally, we focus on  a comparison with experimental results, including tori with either tall or square cross-sections, and propose an interpretation of the experimental observations.

In the next section, we will describe the model and methods used. In section \ref{sec:stst} we will discuss the nature of mean turbulent flows and compare with theoretical steady state predictions. We discuss also the link between Dean vortices and boundary layers.
Then, in section \ref{sec:regimes}, we characterize the different flow regimes and explicit the mean flow power laws. We also point out the influence of the geometry on the nature of the flows. Finally, in section \ref{sec:exp}, we provide a comparison between experimental and numerical results and conclude in section \ref{sec:concl}.

%%%%%%%%%%%%%%%%%%%%%%%%%%%%%%%%%%%%%%%%%%%%%
%%%%%%%%%%%%%%%%%%%%%%%%%%%%%%%%%%%%%%%%%%%%%
\section{Set-up, methods and parameters}\label{sec:set-up}
%%%%%%%%%%%%%%%%%%%%%%%%%%%%%%%%%%%%%%%%%%%%%
%%%%%%%%%%%%%%%%%%%%%%%%%%%%%%%%%%%%%%%%%%%%%

%%%%%%%%%%%%%%%%%%%%%%%%%%%%%%%%%%%%%%%%%%%%%
%%%%%%%%%%%%%%%%%%%%%%%%%%%%%%%%%%%%%%%%%%%%%
\subsection{Geometry and governing equations}
%%%%%%%%%%%%%%%%%%%%%%%%%%%%%%%%%%%%%%%%%%%%%
%%%%%%%%%%%%%%%%%%%%%%%%%%%%%%%%%%%%%%%%%%%%%

\begin{figure}
\begin{center}
\includegraphics[width=8cm]{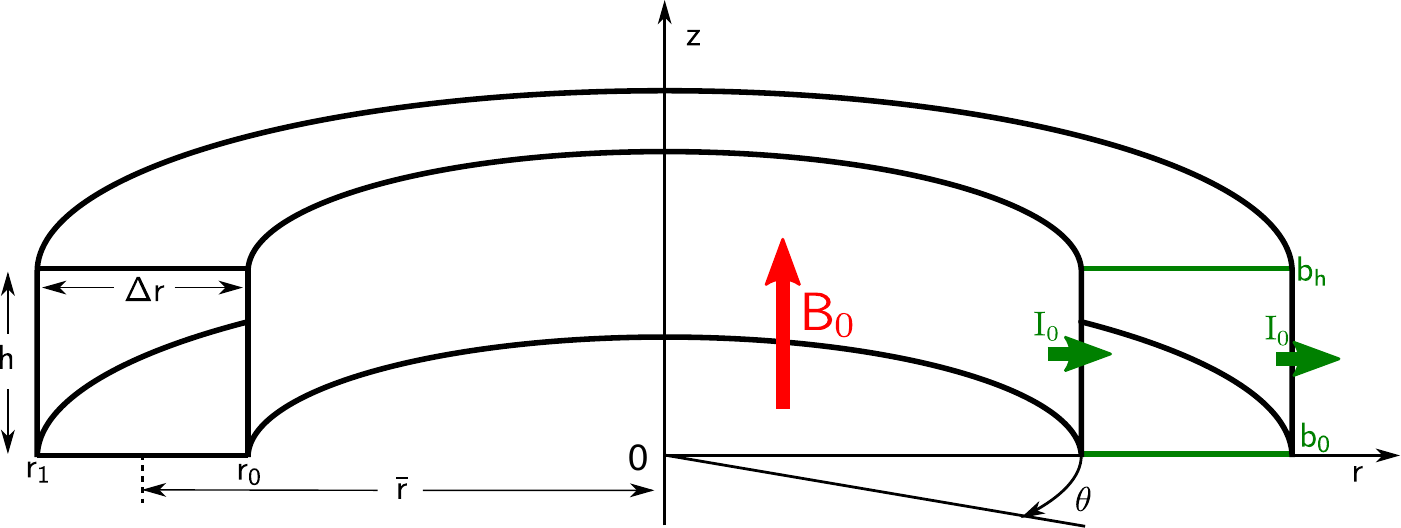}
\caption{ (Color online) Scheme of the annular duct. The axisymmetric 2D simulations presented in this study are displayed on the $(r,z)$ plane.} \label{fig:alb_manip}
\end{center}
\end{figure}

We consider an annular channel of height $h$, inner-radius $r_0$ and outer-radius $r_1$, filled with an electrically conducting fluid and in which a flow is generated -- mainly in the azimuthal ($\theta$) direction --  from the interaction of an imposed axial magnetic field of intensity $B_0$ and an electric current $I_0$ in the radial direction, as sketched in Fig.~\ref{fig:alb_manip}. We will restrict our study to cases where the electric current is injected from the inner radial channel wall. The system is governed by the incompressible magnetohydrodynamic (MHD) equations:
\hskip-1.5cm\begin{eqnarray}
\rho\partial_t \mathbf{V} &=& -\rho\mathbf{V}\cdot\nabla\mathbf{V} -\nabla P +{\bf j}\times\mathbf{B} +\rho\nu\mathbf{\triangle}\mathbf{V}, \label{eq:V}\\
\partial_t \mathbf{B} &=& \mathbf{\nabla}\times\left(\mathbf{V}\times\mathbf{B} \right) + \frac{\eta}{\mu_0}\triangle\mathbf{B}, \label{eq:B} \\
\mathbf{\nabla}\cdot \mathbf{V} &=& \mathbf{0} \label{eq:divV}, \\
\mathbf{\nabla}\cdot \mathbf{B} &=& \mathbf{0} \label{eq:divB},
\end{eqnarray}
where $\mathbf{V}$ is the fluid velocity, $\mathbf{B}$ the magnetic field, ${\bf j}=\mu_0^{-1}\nabla\times {\bf B}$ the current density, $\mu_0$ the magnetic permeability of vacuum, $\rho$ the fluid density, $\eta$ the fluid resistivity and $\nu$ the fluid kinematic viscosity. The magnetic and velocity fields are respectively normalized by the applied axial magnetic field $B_0$ and by the Alfv\`en speed $V_A = B_0/\sqrt{\rho\mu_0}$, while spatial distances are normalized by the channel height $h$. We limit our study to liquid mercury at room temperature. 
One salient feature of the dynamics in MHD devices is the presence of narrow viscoresistive layers at the wall which impact the overall dynamics of the system. The thickness of those layers scale with powers of the Hartmann number (the ratio of Lorentz to viscous forces in Eq.~\eqref{eq:V}), conveniently defined as:
\begin{equation} \label{eq:Ha_RA_LU}
 H\!a=\sqrt{L\!u R_A} =\frac{hB_0}{\sqrt{\rho \eta \nu}}\;.
\end{equation}
where the Lundquist number (the ratio of the magnetic induction to the magnetic diffusion in Eq.~\eqref{eq:B}) and the Alfv\`enic Reynolds number are defined, respectively, as
\begin{equation}
L\!u= \frac{\mu_0 h V_A}{\eta},~~~~R_A = \frac{hV_A}{\nu}.
\end{equation}
Another important dimensionless quantity is the Stuart number 
\begin{equation}
N=\frac{H\!a^2}{Re},    
\end{equation} 
which measures the ratio of electromagnetic forces to inertial forces. Thus, it is linked to the source of current injected into the device. 
The Reynolds number of the flow is given by
\begin{equation}\label{eq:Re}
 Re=\frac{hV_\theta}{\nu}   
\end{equation}
with $V_\theta$ being a characteristic mean azimuthal velocity of the steady state.
 In our simulations $H\!a$ scales from $0.26$ to $827$ and $N$ from $0.068$ to $52$. Steep boundary layers near the walls develop for high values of the Hartmann number. 
Hartmann layers develop in the presence of a magnetic field perpendicular to solid boundaries (e.g. the top and bottom boundaries in Fig.~\ref{fig:alb_manip}) and have a normalized thickness of order $H\!a^{-1}$. Shercliff layers, on the other hand, develop in the presence of a magnetic field parallel to solid boundaries (e.g. the sidewalls in Fig.~\ref{fig:alb_manip}) and have a normalized thickness of order $H\!a^{-1/2}$. Let us emphasize that experimental devices are characterized by different aspect ratio $\epsilon=h/\Delta r$ \cite{Stelzer2015a, Moresco2003, Boisson2012} which impact on the relative width of the layers and thus on the dynamics. Indeed, the layer widths are $\delta_{sh}/\Delta r = \epsilon /\sqrt{H\!a}$ and $\delta_{H\!a}/h = 1/H\!a$.
%
%
%%%%%%%%%%%%%%%%%%%%%%%%%%%%%%%%%
%%%%%%%%%%%%%%%%%%%%%%%%%%%%%%%%%
%
%
Assuming axisymmetry~\cite{Zhao2012} and following the problem formulation derived earlier by Khalzov {\it et al.}~\cite{Khalzov2010}, the velocity and magnetic fields are expressed using stream functions as:
%
%%%%%%%%%%%%%%%%%%%%%%
\begin{eqnarray}
\mathbf{V}/V_A &= &  \frac{u(r,z)}{r}\mathbf{e_\theta} + \frac{1}{r}\nabla w(r,z)\times\mathbf{e_\theta},\label{V_def}\\
\mathbf{B}/B_0 &= & \mathbf{e_z} + \sqrt{\frac{L\!u}{R_A}}\left(\frac{b(r,z)}{r}\mathbf{e_\theta} + \frac{1}{r}\nabla\psi(r,z)\times\mathbf{e_\theta} \right),\label{B_def}
\end{eqnarray}
%%%%%%%%%%%%%%%%%%%%%%%
%
where $r$ and $z$ are respectively the radial and axial coordinates, $u$ is the toroidal angular momentum of the flow, $w$ is the velocity stream function, $b$ is the "angular" toroidal magnetic field and $\psi$ is the toroidal potential vector component. The right hand side quantities in Eq.~(\ref{V_def}-\ref{B_def}) are dimensionless.  

Eq.~(\ref{eq:V}-\ref{eq:divB}) are then recast as:
%
%%%%%%%%%%%%%%%%%%%%%%%%%%%%%%%%%%%%%% 
\begin{eqnarray} \label{eq:model_full}
\partial_t u& \text{=}& \triangle^\ast u \text{+} H\!a\partial_z b \text{+}\frac{R_A}{r}\left\{ u,w \right\} \text{-}\frac{L\!u}{r}\left\{ b,\psi \right\}, \label{eq:model_full1}\\
\partial_t b &\text{=}& \triangle^\ast b \text{+} H\!a\partial_z u \text{+}rL\!u\left\{ \psi,\frac{u}{r^2} \right\} \text{-}rL\!u\left\{ w,\frac{b}{r^2} \right\}, \label{eq:model_full2}\\
\partial_t\triangle^\ast w &\text{=}& \triangle^\ast\triangle^\ast w \text{+} H\!a\partial_z \triangle^\ast \psi \nonumber\\
              & & \text{-}R_A\left( r\left\{ w,\frac{\triangle^\ast w}{r^2} \right\} \text{+}\frac{1}{r^2}\partial_z u^2 \right) \nonumber\\
             & &\text{-}L\!u\left( r\left\{ \psi,\frac{\triangle^\ast \psi}{r^2} \right\} \text{+}\frac{1}{r^2}\partial_z b^2 \right), \label{eq:model_full4}\\
\partial_t \psi &\text{=}& \triangle^\ast \psi \text{+} H\!a\partial_z w \text{+}\frac{L\!u}{r}\left\{ \psi,w \right\}, \label{eq:model_full5}
\end{eqnarray}
%%%%%%%%%%%%%%%%%%%%%%%%%%%%%%%%%%%%%% 
%
where  $\triangle^\ast w$ is the toroidal vorticity, $\triangle^\ast$ is the laplacian in cylindrical coordinates and $\left\{ ., .\right\}$ denotes the Poisson bracket: 
\begin{eqnarray} 
\triangle^\ast u& =& \partial_z^2 u + \partial_r^2 u - \frac{1}{r}\partial_r u, \\
\left\{ u,w \right\} &=&  \partial_r u \partial_z w - \partial_r w \partial_z u.
\end{eqnarray}
%

%%%%%%%%%%%%%%%%%%%%%%%%%%%%%%%%%%%%%%%%%%%%%
%%%%%%%%%%%%%%%%%%%%%%%%%%%%%%%%%%%%%%%%%%%%%
\subsection{Boundary conditions and numerical method}\label{sec:tool}
%%%%%%%%%%%%%%%%%%%%%%%%%%%%%%%%%%%%%%%%%%%%%
%%%%%%%%%%%%%%%%%%%%%%%%%%%%%%%%%%%%%%%%%%%%%

The boundary conditions for the magnetic field $\mathbf{B}$ are fixed by constraints from the wall to the current. Indeed, the normalized current density
%
%%%%%%%%%%%%%%%%%%%%%%%%%%%%%
\begin{equation}\label{eq:j}
\mathbf{j}(r,z) = \sqrt{\frac{L\!u}{R_A r^2}}\left( -\partial_z b(r,z) \mathbf{e_r} -\triangle^\ast \psi(r,z) \mathbf{e_\theta} +\partial_r b(r,z) \mathbf{e_z} \right)
\end{equation}
%%%%%%%%%%%%%%%%%%%%%%%%%%%%
%
is tangential to the top and bottom insulating walls and is perpendicular to side conductive walls. It implies the following boundary conditions on  $b$ and $\psi$: 
\begin{eqnarray}
b(r)|_{z=0,h} &=& b_0,b_h,\label{eq:bound_b}\\
\partial_z b(z)|_{r_0,r_1} &=& 0,0,\\
\psi(r)|_{z=0,h} &=& \psi(z)|_{r=r_0,r_1}=0,\\
\triangle^\ast \psi(z)|_{r=r_0,r_1} &=& 0,\\
\partial_r \triangle^\ast \psi(r)|_{z=0,h} &=& 0.
\end{eqnarray}
$b_h$ and $b_0$ are the values of $b$ respectively at the top and the bottom of the duct. They fix the value of the injected current $I_0$ as: 
\begin{eqnarray}\label{eq:I_b}
I_0 &=& \sqrt{\frac{L\!u}{R_A}}\frac{2\pi}{\mu_0}(b_0 - b_h) \;\;[A]. 
\end{eqnarray}
Let us emphasize that those boundary conditions impose in fact that the total radial current flowing in the duct through any surface $r=\mbox{Const}$ is constant and equals to $I_0$.

No slip boundary conditions for the velocity field at the duct walls impose: 
\begin{eqnarray}
u(z)|_{r=r_0,r_1} &=& u(r)|_{z=0,h}= 0,\\
 w(z)|_{r=r_0,r_1}&=&w(r)|_{z=0,h} =0,\\
\partial_z w(z)|_{r=r_0,r_1} &=& 0,\\
\partial_r w(r)|_{z=0,h} &=& 0.
\end{eqnarray}
The system is integrated using an explicit fourth order {\it Runge-Kutta}  temporal scheme combined with an inhomogeneous finite differences scheme \cite{Sanmiguel2005}. The  inversion of the Laplacians is handled by a {\it successive over-relaxation} algorithm\cite{Press1992} (SOR). The code is parallelized using the Message Passing Interface library. In order to properly resolve the Hartmann and Shercliff layers, the spatial mesh is refined close to the duct walls as:
\begin{eqnarray}
\delta r(r) &= &1 -\exp(K_r(r_0 - r)) -\exp(K_r(r - r_1)) \nonumber\\
& &+ \exp(K_r(r_0 - r_1)), \\
\delta z(z) &= &1 -\exp(-K_zz) -\exp(K_z(z - h)) \nonumber\\
& &+ \exp(-K_z h),
\end{eqnarray}
where $\delta r$ and $\delta z$ are respectively the radial and axial mesh sizes. The parameters $K_{r}$ and $K_{z}$ are adapted according to the device size and the value of $H\!a$. The mesh refinement can exceed a factor of ten at the boundaries for large $H\!a$ values.
In total, we have carried out a few hundreds simulations with typical 200*200 grid-points. Time integration was carried out until statistically steady states were observed for a long enough time interval to converge the statistics.

%%%%%%%%%%%%%%%%%%%%%%%%%%%%%%%%%%%%%%%%%%%%%
%%%%%%%%%%%%%%%%%%%%%%%%%%%%%%%%%%%%%%%%%%%%%
\subsection{Numerical set-up: Experimental control parameters or dimensionless numbers?}
%%%%%%%%%%%%%%%%%%%%%%%%%%%%%%%%%%%%%%%%%%%%%
%%%%%%%%%%%%%%%%%%%%%%%%%%%%%%%%%%%%%%%%%%%%%

The number of parameters to characterize MHD duct flow (geometry, injected current, imposed magnetic field) is substantial, and it is still an open question how to properly identify the different regimes. The characterization of the flows can be presented as a function of either dimensionless numbers or control parameters. When an investigation is presented as a function of dimensionless numbers, it is not straightforward to compare to experimental results, The modification of a single experimental control parameter (current intensity strength of the applied magnetic field) has implications on several dimensionless parameters. For instance, a modification of the strength of the applied magnetic field will simultaneously change the Stuart number, the Hartmann number and the Lundquist number. Despite the elegance of a parameter scan as a function of dimensionless parameters, we have therefore chosen to present our results as a function of the physical control parameters encountered in experiments.

Our approach is therefore to consider the material-properties (density, electrical resistivity and viscosity) of liquid mercury at room temperature and to study the influence of the values of the magnetic field, injected current and of the geometry of the experiment (aspect ratio and radius). The parametric values in the simulations are selected in order to assess and extend previous numerical and experimental studies\cite{Moresco2003,Khalzov2010,Boisson2017}.

In our parameter scans, unless explicitly mentioned, we will consider one particular reference case with experimentally convenient parameters and we will vary one control parameter at a time with respect to this case. This reference case is set with the following parameters: $B_0=0.1$~T, $h=1$~cm, $r_0=4$~cm, $r_1=5$~cm. The injected current is  $I_0=0.5$~A with $b_h = -b_0 = 0.13$. The fluid is mercury ($\rho=13550$~kg/m$^3$, $\nu=1.14\cdot 10^{-7}$~m$^2$/s, $\eta = \mu_0 7.7\cdot 10^{-1}$~$\Omega m$). In our simulations the magnetic field $B$ ranges from $10^{-3}$ to 1~T, the current $I_0$ from 0.1 to 20~A, the height $h$ from 0.25 to 32 cm and the radius $\overline r$  from 1 to 450~cm. 

In order to facilitate the comparison of our results with respect to other studies\cite{Moresco2003,Khalzov2010,Boisson2017}, we present in table 1 the parameter ranges that we have explored in terms of the Hartmann, Reynolds numbers and inertial criterion (see definition in table 1). Note that for our reference case: $H\!a = 25.9$, $Re = 1158$, $L\!u=0.01$, $R_A=67081$, $N=1.73$ and the inertial criterion $B_aN^{-2} = 0.17$. 
%
%%%%%%%%%%%%%%%
\begin{center}
%
%%%%%%%%%%%%%%%%%%%%%%%%%%%%%%%%%
\begin{table}
\begin{tabular}{|c|c|c|c|} 
\hline
 \multirow{2}{*}{} & Hartmann & Reynolds & Inertial criterion \\
                   & number & number &   \\
                           & $h B_0 / \sqrt{\rho \nu}$ & $h V_\theta/\nu$ &  $(2h^2/\bar{r}^2) Re/H\!a^2$ \\
\hline
Ref case with & \multirow{2}{*}{$25.9$} & \multirow{2}{*}{$\left[254\text{-}1.0\cdot 10^4%e^{4}
\right]$} & \multirow{2}{*}{$\left[0.04\text{-}1.5\right]$}\\
$I_0~\in~ \left[ 0.1\text{-}20\right]$~A & & &  \\
\hline
Ref case with  & \multirow{2}{*}{$\left[ 0.26\text{-}258.6\right]$} & \multirow{2}{*}{$\left[ 35\text{-}1766\right]$} & \multirow{2}{*}{$\left[51.8\text{-}2.6\cdot 10^{-3}\right]$}\\
$B_0~\in~ \left[ 0.001\text{-}1\right]$~T & & &  \\
\hline
Ref case with & \multirow{2}{*}{$25.9$} & \multirow{2}{*}{$\left[2870\text{-}13\right]$} & \multirow{2}{*}{$\left[8.4\text{-}1.9\cdot 10^{-7}\right]$}\\
$\bar{r}~\in~ \left[ 1\text{-}450\right]$~cm & & &  \\ 
\hline
Ref case with & \multirow{3}{*}{$\left[ 6.5\text{-}827.7\right]$} & \multirow{3}{*}{$\left[285\text{-}6484\right]$} & \multirow{3}{*}{$\left[0.04\text{-}0.95\right]$} \\
$\Delta r =1$~cm   & & & \\
$h~\in~ \left[ 0.25\text{-}32\right]$~cm & & & \\
\hline
\end{tabular}
\caption{Values of non-dimensional numbers within the range of explored parameters.}
\end{table}
%%%%%%%%%%%%%%%%%%%%%%%%%%%%%%%%%
\end{center}
%
%%%%%%%%%%%%%%%
%
%\clearpage\newpage
%%%%%%%%%%%%%%%%%%%%%%%%%%%%%%%%%%%%%%%%%%%%%
%%%%%%%%%%%%%%%%%%%%%%%%%%%%%%%%%%%%%%%%%%%%%
\section{Asymptotic steady state flows: basic properties and mechanisms}\label{sec:stst}
%%%%%%%%%%%%%%%%%%%%%%%%%%%%%%%%%%%%%%%%%%%%%
%%%%%%%%%%%%%%%%%%%%%%%%%%%%%%%%%%%%%%%%%%%%%
%
Due to the substantial number of control parameters, no comprehensive understanding and consensus exists on the different flow regimes observed in magnetically driven annular duct flow. Before exploring the different possible regimes, we will study the low Lundquist number case treated analytically and numerically\cite{Khalzov2010} in order to benchmark our code and to check if the steady state regime previously obtained does exist when the full time-dependent system is solved. Indeed,  in the previous investigation, the nonlinear dynamics showing the convergence to the steady state was not assessed, but only the time-independent system was solved.

%%%%%%%%%%%%%%%%%%%%%%%%%%%%%%%%%%%%%%%%%%%%%
%%%%%%%%%%%%%%%%%%%%%%%%%%%%%%%%%%%%%%%%%%%%%
\subsection{Low Lundquist number limit: Existence of steady state flows}\label{sec:Khalzov}
%%%%%%%%%%%%%%%%%%%%%%%%%%%%%%%%%%%%%%%%%%%%%
%%%%%%%%%%%%%%%%%%%%%%%%%%%%%%%%%%%%%%%%%%%%%
In earlier work\cite{Khalzov2010} analytical solutions for the toroidal velocity component $u(r,z)$ were obtained in the low magnetic Prandtl number limit, setting $L\!u=0$. They assumed the time independence of the model Eq.~(\ref{eq:model_full1}-\ref{eq:model_full5}). In this limit,  $\psi$ becomes a passive scalar, solution of $\triangle^\ast \psi + Ha\partial_z w = 0$. The fields $w(r,z)$, $b(r,z)$ and $\psi(r,z)$ may then be computed numerically. We present here a direct comparison with their results,  for the parameters $R_A = 4000$, $H\!a=30$ and $L\!u=0$.
%
%%%%%%%%%%%%%%%%%%%%%%%%%%%%%%%%%
\begin{figure}
\begin{center}
\includegraphics[width=8cm]{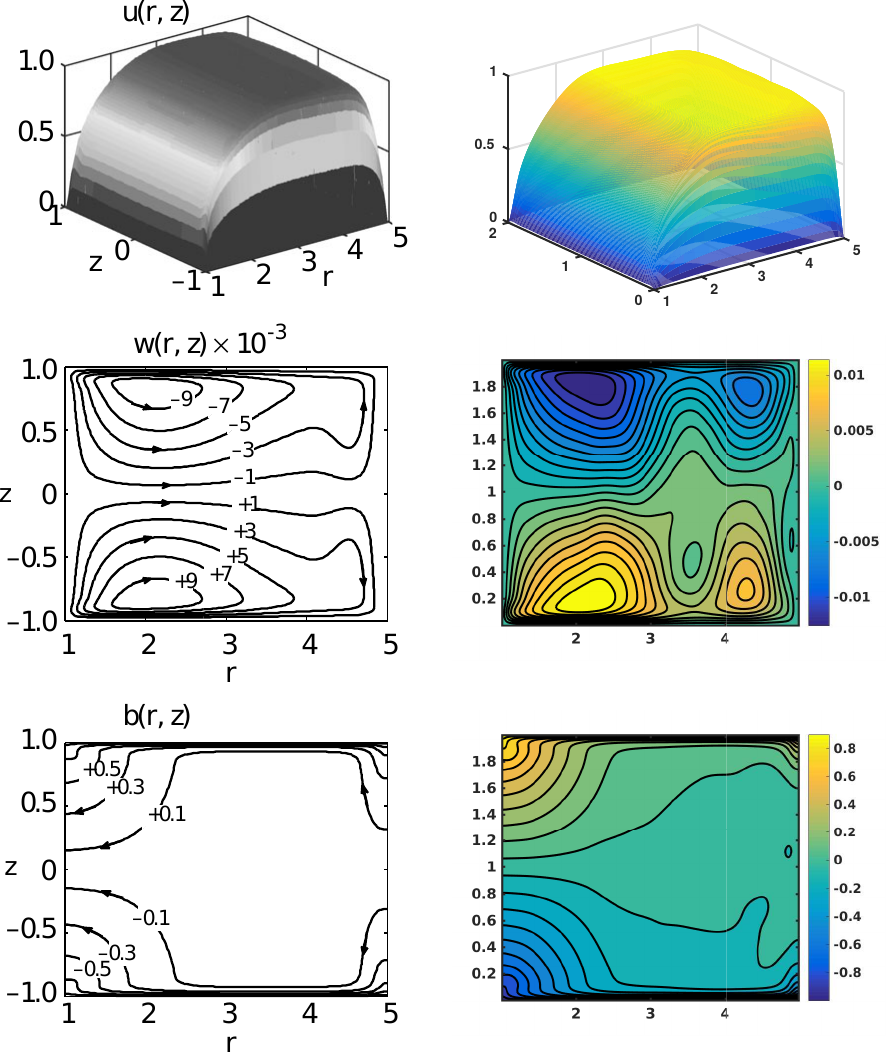}
\caption{ (Color online) Fields $u(r,z)$, $w(r,z)$ and $b(r,z)$ for $R_A = 4000$, $h=2$, $r_0=1$, $r_1=5$ and $H\!a=30$. [Left column]:  Equilibrium solution in the limit $L\!u=0$. [Right column]: Temporal average of the fields during the asymptotic state for the model  Eq.~(\ref{eq:model_full1}-\ref{eq:model_full5})  with $L\!u=0.225$.} \label{fig:Khalsov}
\end{center}
\end{figure}
%%%%%%%%%%%%%%%%%%%%%%%%%%%%%%%%%
%
We have performed a successful benchmark of their results, using a global SOR method in our code which computes the time independent solution for $L\!u=0$. 

\begin{figure}
\begin{center}
\includegraphics[width=8cm]{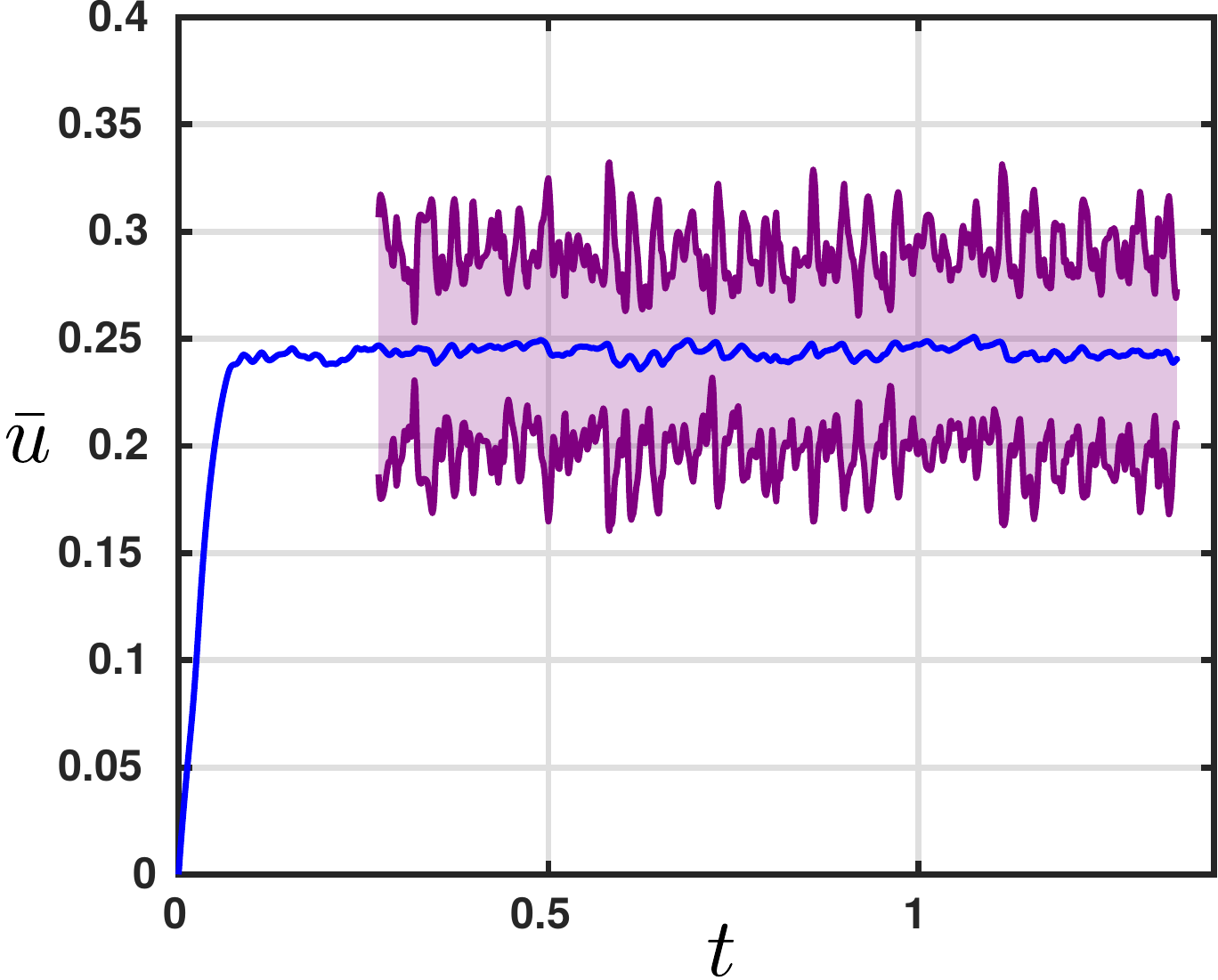}
\caption{ (Color online) Time evolution of $\bar{u}$ the spatial average of $u$, with the model Eq.~(\ref{eq:model_full1}-\ref{eq:model_full5}). Surrounding purple curves are the spatial average of the standard deviation $\delta u$. $R_A = 4000$, $h=2$, $r_0=1$, $r_1=5$, $L\!u=0.225$ and $H\!a=30$.} \label{fig:Khalsov_time}
\end{center}
\end{figure}

Subsequently we compute the time-dependent solution for a a low but nonzero value of the Lundquist number, $L\!u = H\!a^2 / R_A = 0.225$. We observe that the flow does not relax to a static time-independent solution. It turns out that significant temporal fluctuations are observed around a statistically steady state. In  Fig.~\ref{fig:Khalsov_time}, we observe that the amplitude of the fluctuations  of the mean toroidal flow $\delta u$ (purple zone) constitute approximately $40\%$ the value of the mean flow $\bar{u}$. 
However, the time-independent solution\cite{Khalzov2010} corresponds to the time-averaged statistical steady state obtained in our simulations. Fig.~\ref{fig:Khalsov} compares the steady state results obtained by Khalzov {\it et al} \onlinecite{Khalzov2010}. to our time-averaged steady state fields. We note the presence of very narrow Hartmann layers in the vicinity of, and all along the horizontal walls, while the Shercliff layers at the vertical walls are less pronounced, as expected.

%%%%%%%%%%%%%%%%%%%%%%%%%%%%%%%%%
%%%%%%%%%%%%%%%%%%%%%%%%%%%%%%%%%

The study of the underlying instabilities and of the turbulent fluctuations is left for future work. However, this result shows that before those devices can be used to study the MRI instability, one needs to carefully assess the flow properties and in particular the other types of instabilities that might be present in the system. Let us emphasize, however, that the temporal fluctuations do not strongly modify the base-flow, and that the time-averaged turbulent state is very similar to the previously described zero-Lundquist-number stationary flow\cite{Khalzov2010}.

In the remainder of this investigation, we will investigate the time-averaged flow during the statistically steady state. Unless other specified, all quantities denote in the following time-averages. Volume averages will be specified, where needed, by brackets $\left< \cdot \right>$.

%clarify the profile dynamics and to identify the different regimes. 

%%%%%%%%%%%%%%%%%%%%%%%%%%%%%%%%%%%%%%%%%%%%%
%%%%%%%%%%%%%%%%%%%%%%%%%%%%%%%%%%%%%%%%%%%%%
\subsection{Mean flow and origin of the boundary layers \label{sec:MeanFlow}}
\begin{figure}
\begin{center}
\includegraphics[width=1. \linewidth]{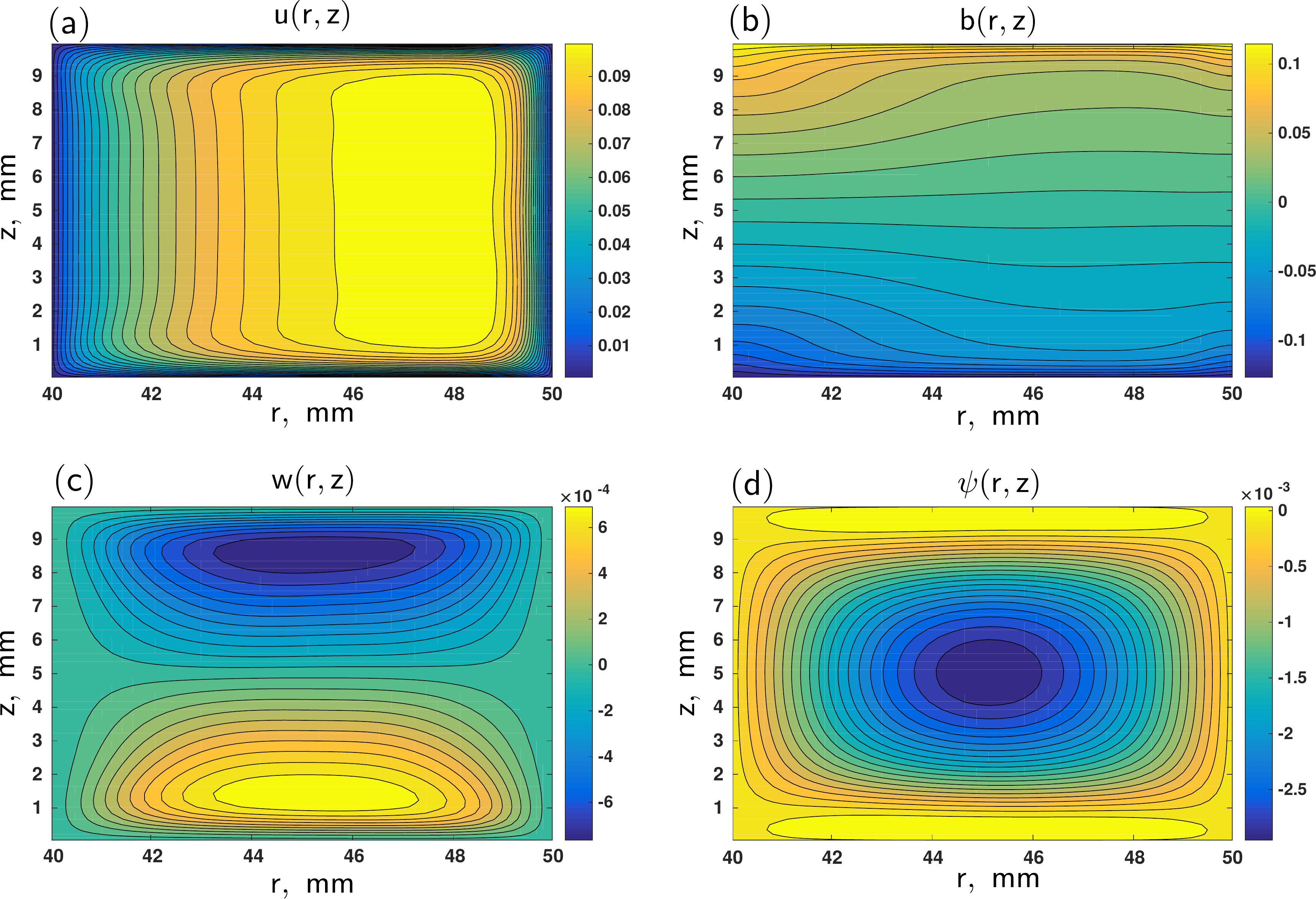}
\caption{(Color online)  (r,z) cross-section of the flow and the current $u(r,z)$, $w(r,z)$, $b(r,z)$ and $\psi(r,z)$ for the reference case: $R_A = 67081$, $L\!u=0.01$ and $H\!a=25.9$. Parameters are $B_0=0.1$~T, $I_0=0.5$~A, $h=1$~cm, $r_0=4$~cm, $r_1=5$~cm. The fluid is mercury ($\rho=13550$~kg/m$^3$, $\nu=1.14\cdot 10^{-7}$~m$^2$/s, $\eta = \mu_0 7.7\cdot 10^{-1}$~$\Omega m$).}\label{fig:flow_scheme}
\end{center}
\end{figure}

%%%%%%%%%%%%%%%%%%%%%%%%%%%%%%%%%%%%%%%%%%%%%
%%%%%%%%%%%%%%%%%%%%%%%%%%%%%%%%%%%%%%%%%%%%%
%
Let us consider the mean flow of the reference case. Compared to the previous section, the normalized parameters are similar but not the geometry. Note that the asymptotic flow is still unsteady but presents weak fluctuations (less than $2\%$) for this reference case. 
The time-averaged flow and magnetic potential in the asymptotic regime are shown in Fig.~\ref{fig:flow_scheme}.
%
%%%%%%%%%%%%%%%%%%%%%%%%%%%%%%%%%
%%%%%%%%%%%%%%%%%%%%%%%%%%%%%%%%%
%
Clearly the boundary layers are much less narrow than in the previous case despite a similar Hartmann number.  The poloidal magnetic field is much lower, as expected when the Lundquist number decreases. The amplitude of the poloidal flow $u$ is similar in both cases.

From these graphs, we can identify the main mechanisms generating the flow, the current and the Hartmann layer: first and at first order, the toroidal flow $V_\theta =u/r$ is generated by the Lorentz force $j_r \times B_0$ along the $\theta$-direction, Fig.~\ref{fig:flow_scheme}a. This dominant flow induces a poloidal magnetic field  for high enough Hartmann numbers ($H\!a\partial_z u$ term in Eq.~\eqref{eq:model_full2}) and creates vertical and radial currents Eq.~\eqref{eq:j}).  The iso-value curves of the toroidal magnetic field $b$, displayed in Fig.~\ref{fig:flow_scheme}b are also electrical current streamlines ${\bf j\times dM}=0 $ ($\psi \ll b$). Clearly, in this case the current radially crosses the core of the channel. The fraction of current expelled toward the insulating horizontal wall, {\it i.e.} the Hartmann layers, is much lower than in the previous case, Fig.~\ref{fig:Khalsov}. It indicates that the Hartmann layer width is not unequivoquely determined by the Hartmann number. 

Simultaneously, the vorticity ${\bf \omega=\nabla\times V}/V_A\approx r^{-1}\nabla u\times {\bf e_\theta}$ ($w \ll u$) is advected by the dominant flow $u$ and thus towards the horizontal wall. This is clear in Fig.~\ref{fig:flow_scheme}a, the isoline of $u$ being the streamlines of the vorticity. 
Toroidal vorticity, and thus the non toroidal component of the flow $r^{-1}\nabla w\times {\bf e_\theta}$ is then generated from the centrifugal force ($r^{-2}R_a\partial_z u^2$ term in Eq.~\eqref{eq:model_full4}) at the layer where $\partial_z\gg\partial_r$, as shown in Fig.~\ref{fig:flow_scheme}c. Isolines of $w$, up to the factor $r^{-1}$, correspond to streamlines of the flow ${\bf V}_{\perp\theta}$. Large convective rolls are observed on both side of the mid-horizontal plane. The latter are generically referred as Dean rolls in the literature~\cite{Dean1928, Potherat2012, Nivedita2017}, and play a part in the generation of the Shercliff layers at the conducting walls. As is clear in Fig.~\ref{fig:Khalsov}. ${\bf V}_{\perp\theta}$ corresponds to the projection of the flow in the $(r,z)$-plane, and will be called in the following the orthogonal flow (orthogonal to the dominant component of the flow). Note that the radial asymmetry of the flow $u(r,z)$ is linked to the Dean vortices which advect and then compress the flow towards the outer wall, and consequently amplify the outer Shercliff layer.  

%%%%%%%%%%%%%%%%%%%%%%%%%%%%%%%%%%%%%%%%%%%%%
%%%%%%%%%%%%%%%%%%%%%%%%%%%%%%%%%%%%%%%%%%%%%
\section{Weak and moderate magnetic Mach number flow regimes}\label{sec:regimes}
%%%%%%%%%%%%%%%%%%%%%%%%%%%%%%%%%%%%%%%%%%%%%
%%%%%%%%%%%%%%%%%%%%%%%%%%%%%%%%%%%%%%%%%%%%%

The features of the time-averaged MHD  annular duct flow are usually analyzed from the balance of the dominant terms of Eq.~\eqref{eq:V}. In previous works\cite{Baylis1971a,Khalzov2010}, it was emphasized that the inertialess regime is reached when  
\begin{equation}
2R_e \ll Ha^2(\bar{r}/h)^2    
\end{equation}
and this defines the the {\it inertial criterion}. The equality is associated with the situation where viscous forces are approximately balanced by the Ampere force near the wall. The opposite inequality corresponds to inertial flows. This estimate involves three dimensionless numbers and it is not clear whether or not some other regimes may exist. Indeed various other force balances are possible. Thus, we first adopt an experimental-like approach by considering a scan varying the injected current $I_0$ at the boundaries and the externally imposed magnetic field $B_0$ in the device in order to identify and/or clarify which regime dominates. We will limit our study to cases where the magnetic Mach number satisfies
$$M_{\mbox{\tiny Ma}}=\frac{V_\theta}{V_A}=\frac{Re}{R_A}\lessapprox 1\;.
$$
which is convenient if one wants to compare with previous works and experiments\citep{Khalzov2010, Boisson2012,Boisson2017, Baylis1971a, Stelzer2015b}. Note however that the experimental work of Moresco and Alboussière  \citep{Moresco2003} addresses the issue of large magnetic Mach numbers, and we will come back to this point later. We fix the dimensions of the geometry in this section to the reference case values.%
%%%%%%%%%%%%%%%%%%%%%%%%%%%%%%%%%
\begin{figure}
\begin{center}
\includegraphics[width=0.90 \linewidth]{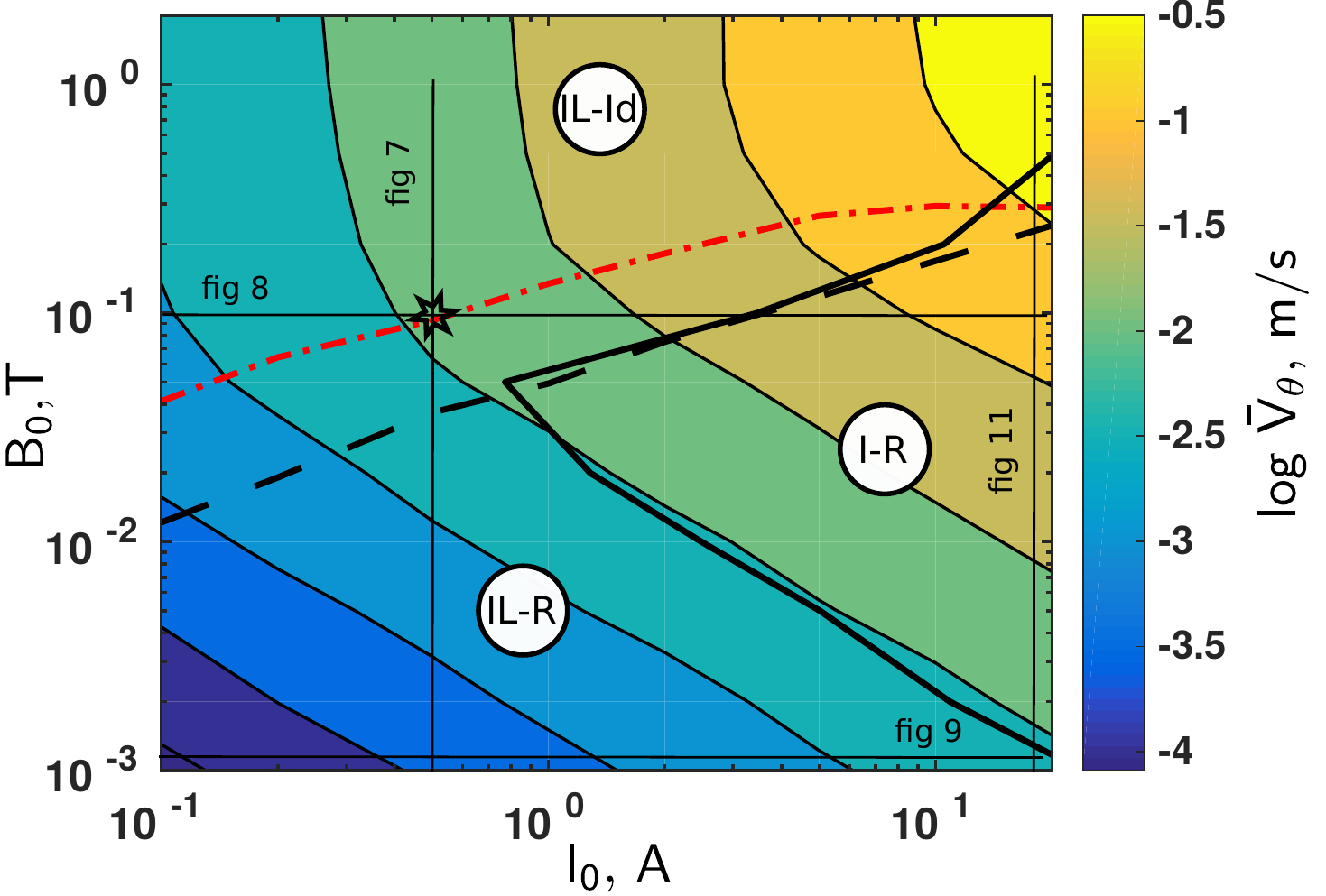}
\caption{ (Color online) Mean toroidal flow versus $B_0$ and $I_0$. Dashed black line corresponds to the critical ratio for the inertial criterion $2h^2Re/H\!a^2\bar{r}^2=1$. Black full line is the numeric estimate of $\left<\|(V\cdot\nabla)V\|\right>/ \left<\|\nu \triangle V \|\right >=0.7$. The red dot-dashed line is the numeric estimate of the induction rate $\tau$, Eq.~\eqref{eq:tau}. There are tree clear regimes: \textsc{Il--Id} the inertialess regime, \textsc{I--R} the inertial regime and  \textsc{Il--R} the weak regime.}\label{fig:regime}
\end{center}
\end{figure}
%%%%%%%%%%%%%%%%%%%%%%%%%%%%%%%%%
%
%
%%%%%%%%%%%%%%%%%%%%%%%%%%%%%%%%%%%%%%%%%%%%%
%%%%%%%%%%%%%%%%%%%%%%%%%%%%%%%%%%%%%%%%%%%%%
\subsection{Characterization of flow regimes as a function of $I_0$ and $B_0$}\label{sec: Regimes identification}
%%%%%%%%%%%%%%%%%%%%%%%%%%%%%%%%%%%%%%%%%%%%%
%%%%%%%%%%%%%%%%%%%%%%%%%%%%%%%%%%%%%%%%%%%%%
%
Computation of the mean toroidal speeds for various values of $B_0$ and $I_0$ are reported in Fig.~\ref{fig:regime}. Regimes are identified by a comparison between the advection and the diffusion term in the toroidal momentum equation Eq.~\eqref{eq:V}. This comparison lead to the inertial criterion limit $2R_e = Ha^2(\bar{r}/h)^2$ which corresponds to the dashed black line in the figure. 

Fig.~\ref{fig:regime} shows two trends for the toroidal velocity, roughly separated by this dashed black line. At high magnetic field $B_0$, the amplitude of the flow does not depend on $B_0$ while for low magnetic field the isovalues indicate that $ V_\theta\propto \left(B_0I_0\right)^\alpha$  where the constant $\alpha>0$ will be specified below.

\begin{figure*}
\begin{center}
\includegraphics[height=2.5cm]{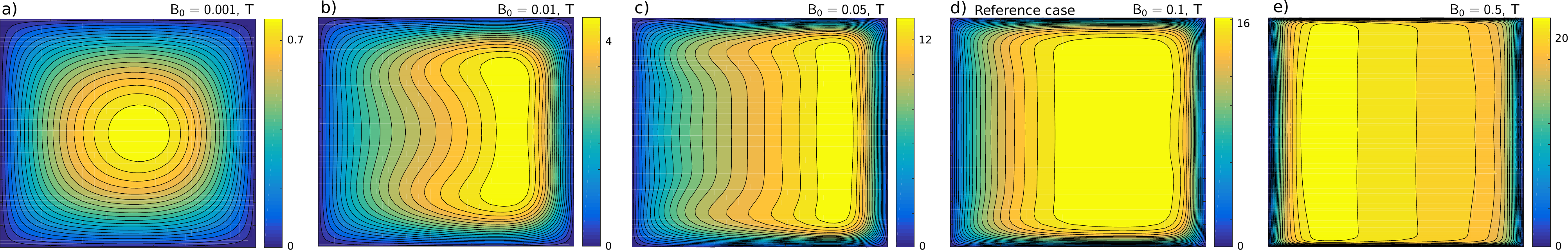}
\caption{ (Color online) Mean toroidal flow velocity snapshots for various magnetic fields $B_0$. Other parameters are reference case ones.}\label{fig:flow_regime_toroidal}
\end{center}
\end{figure*}

Nevertheless, the inertial criterion is not appropriate in the limit of  low current and/or low  magnetic field because the boundary layer widths become a large fraction of the lengths of the device. This is expected from the estimate of the width of the layers and clearly observed in Fig.~\ref{fig:flow_regime_toroidal}a. A more direct estimate of the influence of inertial with respect to viscous effects is the ratio
\begin{equation}
   \frac{ \left<\|(V\cdot\nabla)V\|\right>}{ \left<\|\nu \triangle V \|\right >},
\end{equation}
where the brackets $\left<.\right>$  indicate a volume average. This ratio is a Reynolds number which gives a better estimate than (\ref{eq:Re})  of the influence of inertial effects on the flow.

It is convenient to assess if the regime is inertialess or not for any $I_0$, $B_0$ values. Indeed, it includes all the various terms of the advection and diffusion. $\left<\|(V\cdot\nabla)V\|\right>/ \left<\|\nu \triangle V \|\right >=0.7$ correspond to the black solid line in Fig.~\ref{fig:regime}. For high $B_0$ or $I_0$ values, we obtain that our ratio matches the inertial criteria(the dashed and solid black lines overlap). The inertial regime correspond to the zone on the right of the black solid line in the $(I_0,B_0)$ space, see Fig.~\ref{fig:regime}. 
For low $B_0$ or $I_0$ values, there is another inertialess area. It is not captured by the inertial criterion since, for low $B_0$ or $I_0$, there are no more boundary layers. 
%
%
%%%%%%%%%%%%%%%%%%%%%%%%%%%%%%%%%
%%%%%%%%%%%%%%%%%%%%%%%%%%%%%%%%%
%
%
 We see thus that the two trends in this figure in terms of toroidal flow amplitude are not univoquely discriminated by the usual inertial criterion, but a more refined measure of the Reynolds number is needed.

The dominant terms balanced in the Ohm's law should  discriminate  inertialess regime(s). Indeed, when the induction is important, which happens with a high magnetic field, in other words  when $U\times B_0\gg \eta J$, the toroidal magnetic field is expelled from the center of the channel to the top and bottom boundaries. This mechanism enhances the Ampere force near the wall and thus generates Hartmann boundary layers and stimulates the emergence of the (ideal) inertialess regime. Concomitantly, it flattens also the toroidal magnetic field or equivalently depletes the radial current in the core. If follows that the existence of this inertialess regime is constrained by the necessary condition that radial current has no vertical gradient in the core: 
\begin{eqnarray*}
-\sqrt{\frac{R_A}{L\!u}}\bar{r} J_r(\bar{r},h/2)&=&\partial_z b(\bar{r},h/2)\approx 0\\
& & \mbox{ for inertialess ideal regime.}
\end{eqnarray*}
The opposite case corresponds to a balance between the resistive term and the radial potential in the Ohm's law which is then said resistive ($U\times B_0\ll \eta J$). Then, the injected current is present all over the volume and depend only on $r$. Accordingly, $b \propto B_\theta /r$ becomes a linear function of height. In agreement with the fixed boundary conditions imposed for $b$, we can estimate the current in the middle of the section as:
\begin{eqnarray}
\partial_z b(\bar{r},h/2)\approx 
\frac{b_0-b_h}{2} \mbox{ for resistive regime.}
\label{Eq:resistive regime}
\end{eqnarray}
This help us to build a criterion based on the value of the current in the center of the channel. Indeed, the induction ratio
\begin{equation}
\tau=2\frac{\partial_z b(\bar{r},h/2)}{b_0-b_h} \label{eq:tau}
\end{equation}
is a number in the range $[0,1]$: it goes from 0 for ideal Ohm's law to 1 for resistive Ohm's law. Arbitrarily, the transition between the two regimes can be associated to the ratio value  $\tau=0.5$ and is plotted in Fig.~\ref{fig:regime} using a red dot-dashed line. We observe that the line $\tau=0.5$ corresponds well to the zone of transition in between the two flow behaviors in terms of $(I_0,B_0)$-parametric dependence. $\tau>0.5$ defines the inertialess ideal regime. The inertialess resistive regime is included in $\tau<0.5$. If we define the inertialess resistive regime as $\tau\ge 0.9$, than it corresponds to the zone below the dashed black line, which shows the dependence of this criterion on the exact value of $\tau$.  Henceforward, we will label these regimes respectively \textsc{Il--Id} and \textsc{Il--R} regimes. In Fig.~\ref{fig:regime}, the inertial regime is in fact a resistive inertial regime that we label \textsc{I--R}. Note that for higher current and/or magnetic field, we should reach a fourth regime on the right of the graph, an inertial ideal regime.

Let us specify that the \textsc{Il--Id} regime corresponds to the first regime historically studied with the experiments and calculations of Baylis and Hunt \cite{Baylis1971a, Baylis1971b}. The \textsc{I--R} regime was also described by these authors \cite{Baylis1971a, Baylis1971b} and further investigated more recently\cite{Khalzov2010}. The \textsc{Il--R} regime has not yet been investigated, as far as we are aware of. Concerning the regimes reached in the experimental work of  Moresco and Alboussière  \citep{Moresco2003}, the regimes will be discussed in section \ref{sec:exp_square}.

%%%%%%%%%%%%%%%%%%%%%%%%%%%%%%%%%%%%%%%%%%%%%
%%%%%%%%%%%%%%%%%%%%%%%%%%%%%%%%%%%%%%%%%%%%%
\subsection{Scaling laws}
%%%%%%%%%%%%%%%%%%%%%%%%%%%%%%%%%%%%%%%%%%%%%
%%%%%%%%%%%%%%%%%%%%%%%%%%%%%%%%%%%%%%%%%%%%%
%
An advantage of our reference case is that it is  close to the limit of the three regimes such that scans from this case in current and magnetic field sweep the three regimes (see Fig.~\ref{fig:regime}). In this section we aim to obtain and/or verify the scaling laws which characterize the different regimes. Figs.~\ref{fig:Scan_Vt_ref_B} and \ref{fig:Scan_Vt_ref_I} show respectively the dependence of $V_\theta$ on $B_0$ and $I_0$.  the reference case corresponds to the black star. The blue curves are the numerical results. We compare now the numerical results with theoretical scaling laws in the rest of this section.
%
%%%%%%%%%%%%%%%%%%%%%%%%%%%%%%%%%%%%%%%%%%%%%
%%%%%%%%%%%%%%%%%%%%%%%%%%%%%%%%%%%%%%%%%%%%%
\subsubsection{The Inertialess Ideal regime (\textsc{Il--Id}).}
\begin{figure}
\begin{center}
\includegraphics[width=0.90 \linewidth]{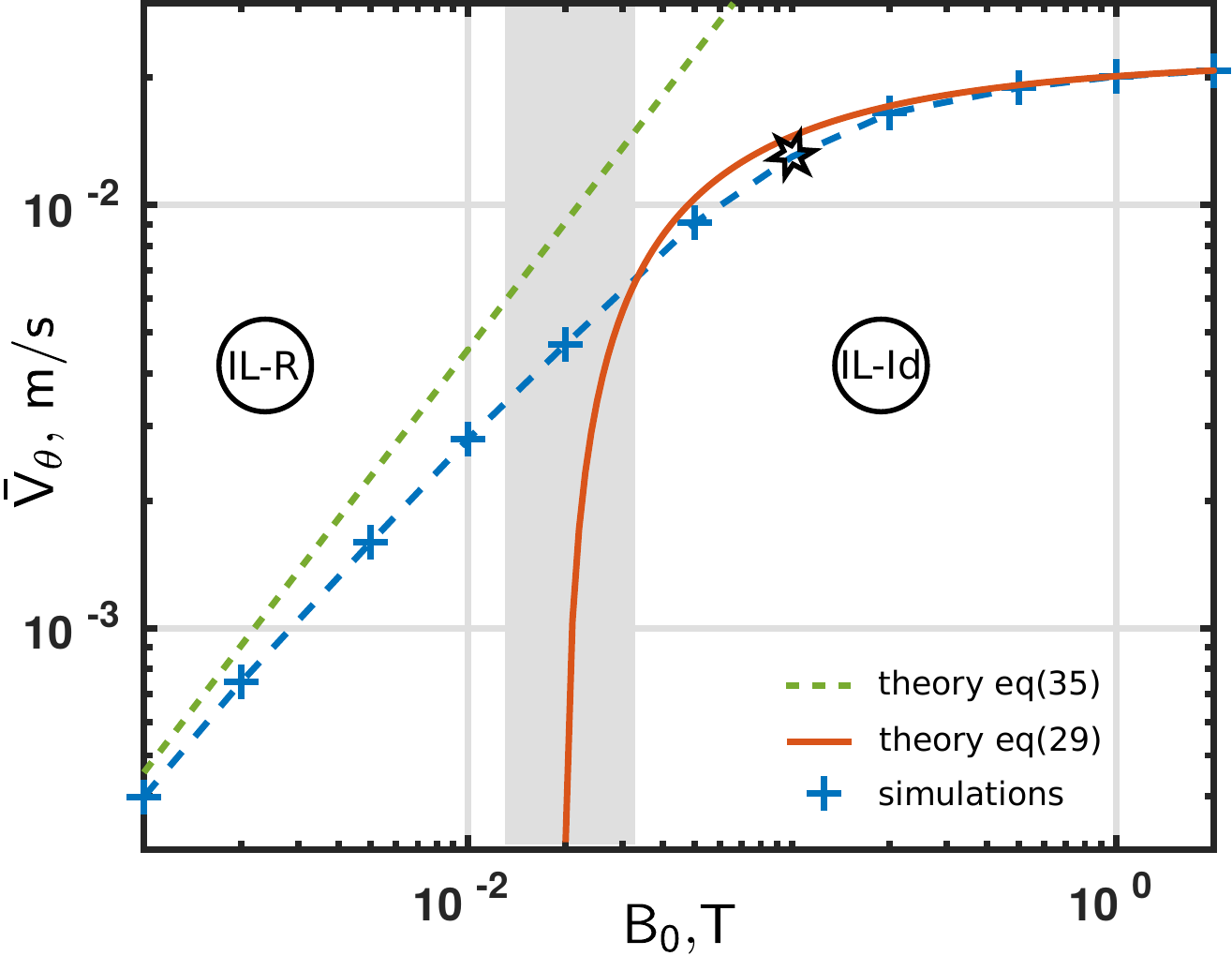}
\caption{(Color online) Mean toroidal flow versus imposed magnetic field. Other parameters are reference case parameters. 
The grey zone separate the limit of validity of the \textsc{Il--Id}  and \textsc{Il--R} regime. Orange Line: Theoretical \textsc{Il--Id} regime.
Green line: Theoretical \textsc{Il--R} regime.} \label{fig:Scan_Vt_ref_B}
\end{center}
\end{figure}

\begin{figure}
\begin{center}
\includegraphics[width=0.90 \linewidth]{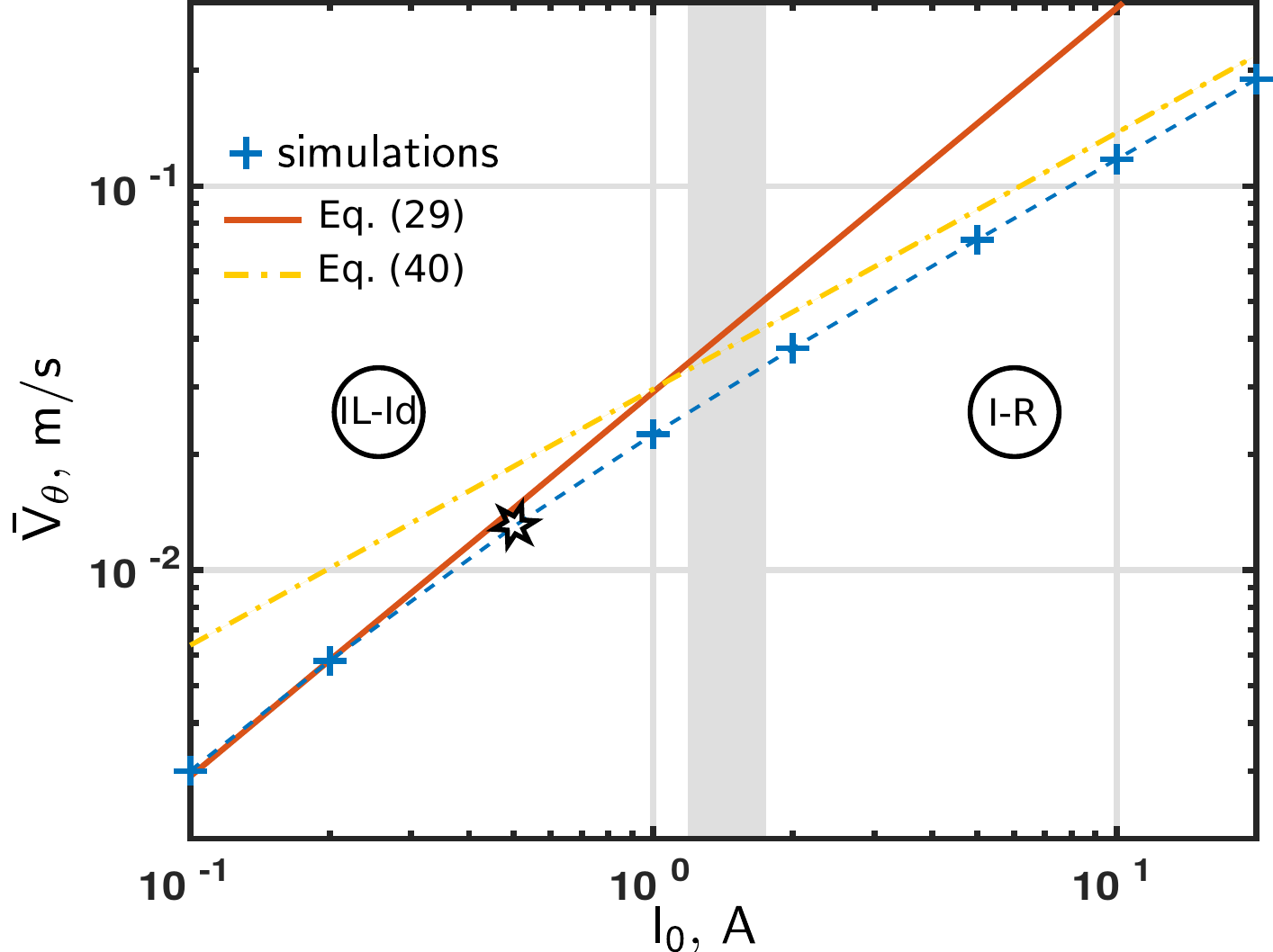}
\caption{ (Color online) Mean toroidal flow versus imposed current. Orange Line: Theoretical mean $V_\theta$ for \textsc{Il--Id} regime, Eq.~\eqref{eq:V_0Baylis}.  Yellow dot-dashed line: Theoretical mean $V_\theta$ for \textsc{I--R} regime, Eq.~\eqref{eq:v2_final}.
The grey zone separate the limit of validity of the \textsc{Il--Id}  and \textsc{I--R} regime.} \label{fig:Scan_Vt_ref_I}
\end{center}
\end{figure}

%%%%%%%%%%%%%%%%%%%%%%%%%%%%%%%%%%%%%%%%%%%%%
%%%%%%%%%%%%%%%%%%%%%%%%%%%%%%%%%%%%%%%%%%%%%
%
We recall here the results of Baylis and Hunt \cite{Baylis1971a,Baylis1971b}. They carry out an experimental and analytical comparison of the main toroidal flow in the inertialess regime. They simplify the system by considering the limit in which perpendicular fluctuations to the toroidal fields can be neglected: $w=\psi=0$. In our notations, they obtained the following solution for the toroidal flow $V_\theta$:

\begin{equation}
V_\theta^{\mbox{\tiny IL-Id}} = \frac{I_0\ln(r_1/r_0)}{4\pi (r_1-r_0)}\sqrt{\frac{\eta}{\rho\nu}} \times \label{eq:V_0Baylis}
\end{equation}
\begin{equation}
 \left(1  - \frac{2}{Ha} - \frac{0.956\sqrt{2}h\ln(r_0/r_1) }{2\sqrt{Ha}}\left(\frac{1}{r_1} + \frac{1}{r_0} \right) \right) \; [m/s]. \notag
\end{equation}
The last two terms on the right hand side are corrections for radially narrow ducts where the Shercliff layer $\delta_{Sh}$ becomes comparable to $\bar{r}$ or, equivalently,  when the magnetic field is not strong enough. 
It is enlightening and straightforward to extract the dominant term of this formula. Indeed, in such a regime, as already discussed, the radial current is expelled towards the Hartmann layers. Considering that all the current concentrates uniformly in these layers of total axial width $2h/H\!a$ and localized at $\bar{r}$, an estimate of the current density in the layers is
\begin{equation}
j\simeq \frac{I_0 H\!a}{2\pi \bar{r} 2 h}\;.\label{eq:j_inless}
\end{equation}
The toroidal velocity is then obtained by the balance between the diffusion term and the lorentz force in the Hartmann layers:
\begin{equation}
V_\theta^{\mbox{\tiny IL-Id}}\propto \frac{I_0}{4\pi \bar{r}}\sqrt{\frac{\eta}{\rho\nu}}
\propto \frac{I_0\ln(r_1/r_0)}{4\pi (r_1-r_0)}\sqrt{\frac{\eta}{\rho\nu}}
 \; [m/s],
\end{equation}
where we have used that $r_0\lessapprox r_1$. This expression is consistent with Eq.~\eqref{eq:V_0Baylis}. Note that it gives the scaling law for the current and the geometric factor but not for the magnetic field which appears in higher order terms.
Eq.~\eqref{eq:V_0Baylis}  predicts $V_\theta$ in this regime, and is displayed as a solid orange line in Figs.~\ref{fig:Scan_Vt_ref_B} and \ref{fig:Scan_Vt_ref_I}.
We find that numerical and theoretical results agree well for $B_0>0.03$[T]. This limit corresponds to the end of validity of this regime, as can be observed in Fig.~\ref{fig:regime}. For lower magnetic field strength, Eq.~\eqref{eq:V_0Baylis} does not agree with the numerical results
%
%%%%%%%%%%%%%%%%%%%%%%%%%%%%%%%%%
%%%%%%%%%%%%%%%%%%%%%%%%%%%%%%%%%
%
%%%%%%%%%%%%%%%%%%%%%%%%%%%%%%%%%
%%%%%%%%%%%%%%%%%%%%%%%%%%%%%%%%%
%
%%%%%%%%%%%%%%%%%%%%%%%%%%%%%%%%%%%%%%%%%%%%%
%%%%%%%%%%%%%%%%%%%%%%%%%%%%%%%%%%%%%%%%%%%%%
\subsubsection{The Inertialess Resistive regime (\textsc{Il--R}).}
\begin{figure}
\begin{center}
\includegraphics[width=0.90 \linewidth]{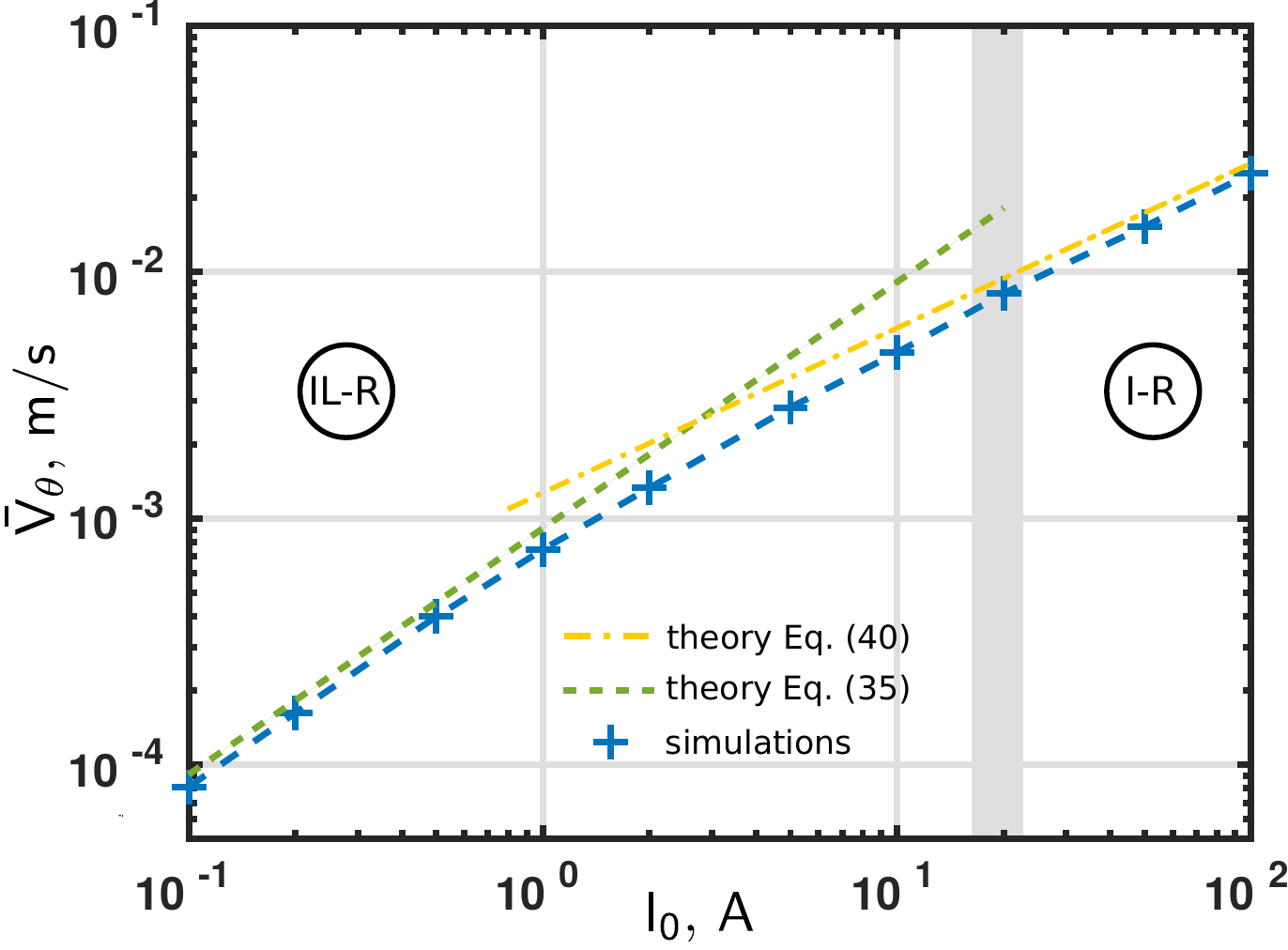}
\caption{  (Color online) Mean toroidal flow for reference case with $B_0$ = 1~mT and with various current values, obtained from simulation (blue crosses). The scan crosses the Weak regime \textsc{Il--R} to the Inertial regime \textsc{I--R} around $I_0$=20~A. The theoretical value Eq.~\eqref{eq:Vt_square_IL_R} in regime \textsc{Il--R} is the green line. The theoretical value Eq.~\eqref{eq:v2_final} in regime  \textsc{I--R} is the yellow dot-dashed line.} \label{fig:Scan_Vt_ref_I_lowB}
\end{center}
\end{figure}

%%%%%%%%%%%%%%%%%%%%%%%%%%%%%%%%%%%%%%%%%%%%%
%%%%%%%%%%%%%%%%%%%%%%%%%%%%%%%%%%%%%%%%%%%%%
%
In the inertialess resistive (Il--R) regime, the advection term is negligible in Eq.~\eqref{eq:V}, as is the induction in Ohm's law, Eq.~\eqref{eq:B}. The radial current density is constant over the duct in a first approximation, see Eq.~\eqref{Eq:resistive regime} (neglecting the cylindrical effect, which is valid when $\bar{r}\gg\Delta r$),
\begin{equation}
j =  \frac{I_0}{2\pi r h}\simeq \frac{I_0}{2\pi \bar{r} h}.\label{eq:j_resistiveregime}
\end{equation}
The steady state of Eq.~\eqref{eq:V} correspond to the balance between the Lorentz force and the viscous term: 
${ j_r B_0\simeq -\rho\nu\Delta V_\theta }$.
The estimate of viscoresistive layer widths in power of $H\!a$ is not valid. Indeed, the width of the layers becomes comparable to the channel sizes ($h$ and/or $\Delta r$). Thus, one obtains:
\begin{equation}
\Delta V_\theta(r,z) = -\frac{I_0B_0}{2\pi\bar{r}h\rho\nu}. \label{eq:Vt_reg3_1}
\end{equation}
Fig.~\ref{fig:flow_regime_toroidal}a shows the numerical solution of $V_\theta$ and its circular shape. It confirms the validity of the assumption leading to Eq.~\eqref{eq:Vt_reg3_1}. With a circular section $\tilde{r} = \sqrt{(r-\bar{r})^2 + z^2}$, one can express the toroidal flow profile by solving Eq.~\eqref{eq:Vt_reg3_1}:
\begin{equation}
V_\theta(\tilde{r}) = \frac{I_0B_0}{8\pi\bar{r}h\rho\nu} \left( \tilde{r}_e^2 - \tilde{r}^2 \right),
\end{equation}
with $\tilde{r}_e$ the effective radius of the channel. The boundary conditions are $V_\theta^{\prime}(\tilde{r}=0)=0$ and $V_\theta(\tilde{r}_e=0)=0$. %Supposing that 
The channel section surface is constant %with a circular or a square section, 
with $\tilde{r}_e = \sqrt{\Delta r h/\pi}$. One obtains for the mean toroidal flow in the \textsc{Il--Id} regime, after a volume integration:
\begin{equation}
V_\theta^{\mbox{\tiny IL-R}} = \frac{I_0B_0 \Delta r}{16\pi^2\bar{r}\rho\nu} \; [m/s].\label{eq:Vt_square_IL_R} 
\end{equation}
This calculation is confirmed in Fig.~\ref{fig:Scan_Vt_ref_B}, where $V_\theta^{\mbox{\tiny IL-R}}$ is shown in green. To test the robustness of the result, another scan is proposed where the value of the current is varied for a given low magnetic field of 1~mT, see Fig.~\ref{fig:Scan_Vt_ref_I_lowB}. In this case, the inertial regime starts for current values higher than 20~A. The good agreement of the numerical results and the predicted values of $V_\theta^{\mbox{\tiny IL-R}}$, confirm the validity of the scaling law for low currents. Note that this regime is also easily obtained using conductive aqueous solutions of CuSO$_4$ with $\sigma = 0.75$~S/m \cite{Digilov2007, Suslov2017}.
%
%%%%%%%%%%%%%%%%%%%%%%%%%%%%%%%%%
%%%%%%%%%%%%%%%%%%%%%%%%%%%%%%%%%
%
%
%%%%%%%%%%%%%%%%%%%%%%%%%%%%%%%%%%%%%%%%%%%%%
%%%%%%%%%%%%%%%%%%%%%%%%%%%%%%%%%%%%%%%%%%%%%
\subsubsection{The Inertial Resistive regime (\textsc{I--R})}
\begin{figure}
\begin{center}
\includegraphics[width=0.90 \linewidth]{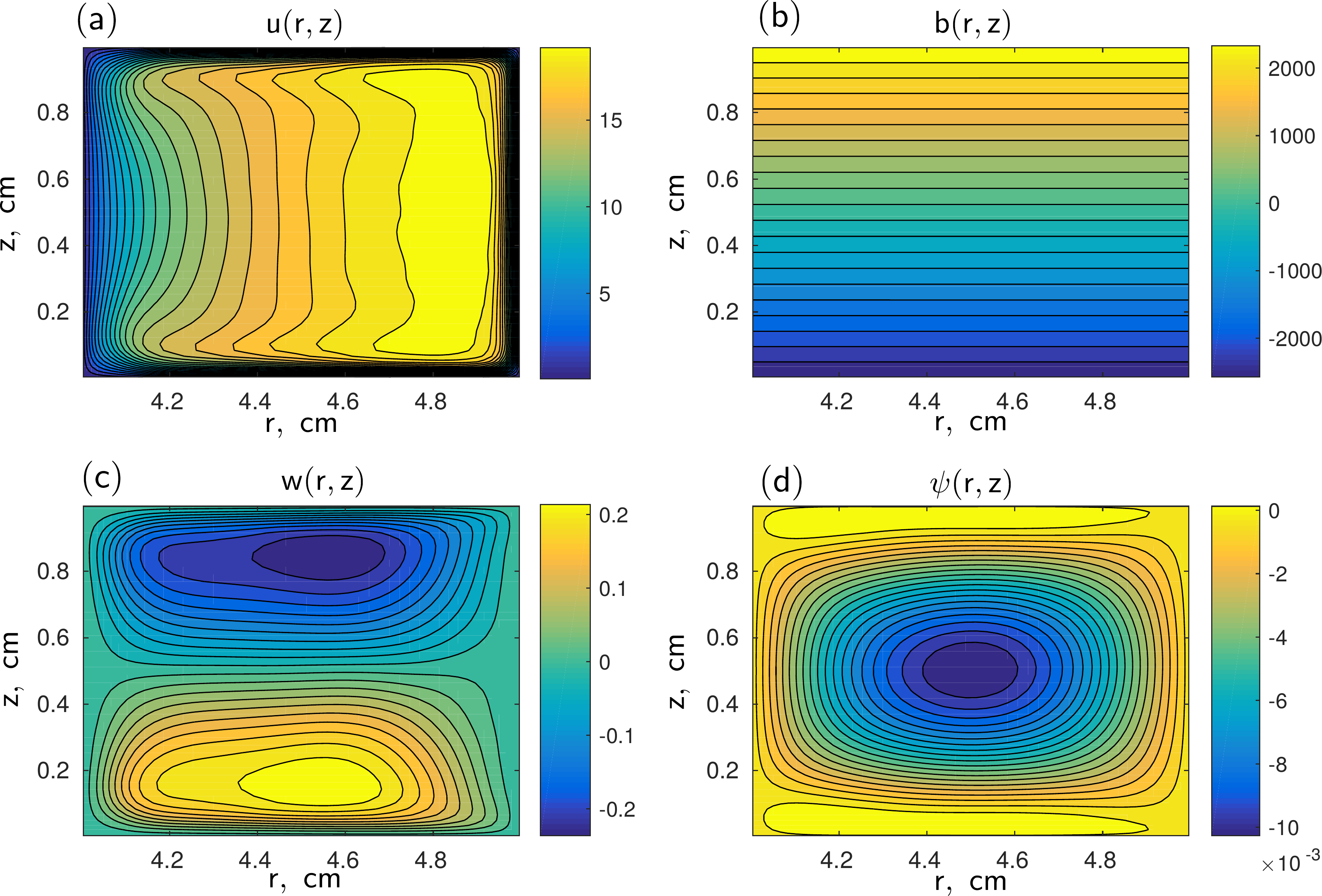}
\caption{  (Color online)  (r,z) cross-section of the flow and the current $u(r,z)$, $w(r,z)$, $b(r,z)$ and $\psi(r,z)$ for the reference case with $B_0 = 1$~mT and $I_0 = 100$~A in inertial regime ($R_A = 672$, $L\!u=10^{-4}$ and $H\!a=0.26$).} \label{fig:Innertialrun}
\end{center}
\end{figure}

%%%%%%%%%%%%%%%%%%%%%%%%%%%%%%%%%%%%%%%%%%%%%
%%%%%%%%%%%%%%%%%%%%%%%%%%%%%%%%%%%%%%%%%%%%%
%
%
In the inertial resistive regime, $V\times B$ can be neglected in Ohm's law. In the core, the current is uniformly distributed in toroidal sections. Thus,  Eq.~\eqref{eq:j_resistiveregime} is valid. The momentum equation in the toroidal direction is characterized by a balance between the advection term and the Lorentz force.
The part of the convective term which balances the Lorentz force in the core,  Eq.~\eqref{eq:V},   is $V_r \partial_r V_\theta$ since
$$\frac{\mathcal{O}(V_z \partial_z V_\theta)}{\mathcal{O}(V_r \partial_r V_\theta)}\approx \frac{\partial_r\!w \partial_z\!u}{\max(\partial_z\!w \partial_ru,u\partial_z\!w /r)}\ll 1\;.$$
Indeed, the presence of Dean's rolls insures $\mathcal{O}(\partial_r\!w) \simeq \mathcal{O}(\partial_z\!w)$, see Fig.~\ref{fig:Innertialrun}c %\ADD{Khalzov c'est bien le régime inertiel, là où ces rouleaux sont forts?}
The field $u$ has almost no $z$-gradient in this regime, see Fig.~\ref{fig:Innertialrun}a. Thus, we have the ordering $|\partial_z\!u |\ll |\partial_r\!u|$. Moreover, the latter figure indicates that the characteristic length of the r-gradient is $\Delta r$ which is smaller than $\bar{r}$. It follows that, as already discussed\cite{Khalzov2010}, the dominant convective term is $r^{-2}u\partial_z\!w=V_rV_\theta/r$
which equals the Coriolis acceleration linked to the curvature of the channel. Thus, in the core of the device 
%
%%%%%%%%%%%%%%%%%%%%%%%%%%%%%%%%%
%%%%%%%%%%%%%%%%%%%%%%%%%%%%%%%%%
%
\begin{equation}
\rho\frac{V_r V_\theta}{\Delta r } \simeq \frac{I_0 B_0 }{2\pi \bar{r} h }.\label{eq:v2_1}
\end{equation}
This advection term includes $V_r$ and involves thus the radial momentum equation balance. The drive of this equation is the curvature force, 
the $V_\theta^2/\bar{r}$ term which arises explicitly in cylindrical coordinates. 

\begin{figure}
\begin{center}
\includegraphics[width=0.90 \linewidth]{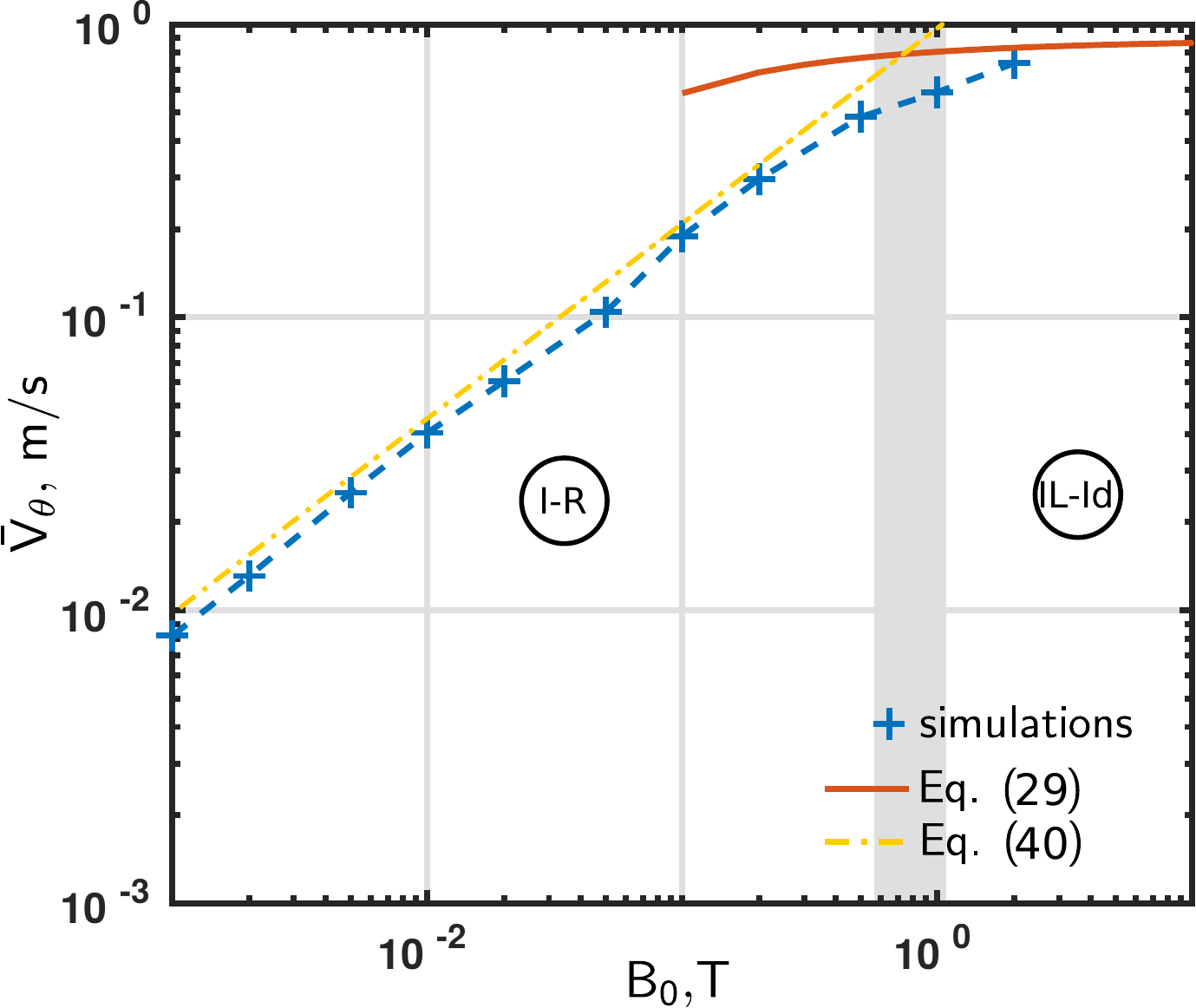}
\caption{  (Color online) Mean toroidal flow for reference case with $I_0$ = 20~A and with various magnetic field values, obtained from simulation (blue crosses). The scan crosses the Inertial regime \textsc{I--R} to the Inertialess regime \textsc{Il--Id} around $B_0$=1~T. The theoretical value Eq.~\eqref{eq:V_0Baylis} in regime \textsc{Il--Id} is the orange line. The theoretical value Eq.~\eqref{eq:v2_final} in regime  \textsc{I--R} is the yellow dot-dashed line.}\label{fig:Scan_Vt_ref_B_lowI}
\end{center}
\end{figure}

We now introduce the notations $V_r^{-}$ and $V_\theta^{-}$ to denote the order of magnitude of the  the radial and toroidal flows, respectively, in the viscous layers.  Large convective rolls (or Dean vortices) connect the radial flow velocity $V_r$ in the core to the one, $V_r^-$, in the horizontal viscous layers. The latter are generated by the strong axial gradient of the toroidal velocity along the wall. They are also bounded by the Dean vortices which advect the fluid up and down from the center. Those rolls have a maximal extension factor $f=1/2$ of the total height.  Of course, as can be seen in Fig.~\ref{fig:Innertialrun}c, they are compressed toward the up and down wall. Thus, an estimate of the viscous layer width between the core of the vortex and the wall is $\delta_\nu = \sqrt{2\pi \bar{r}\nu / V_\theta^-}$. In this zone, the toroidal velocity profile decreases linearly as expected in viscous layers and this is also what we observe in our simulations. The measure of $\delta_\nu$ corresponds to the distance between the wall and the core of the vortex.
%The notations $V_r^{-}$ and $V_\theta^{-}$ stand respectively for the radial and toroidal flows in the viscous layers. 
Because of the linear decrease of  $V_\theta$ in between the walls and Dean vortices, we estimate
\begin{equation}
V_\theta^-\simeq \frac{V_\theta}{2}\;\label{eq:vthetam}
\end{equation}
We emphasize that the toroidal velocity is laminar outside the viscous layer and has almost no axial gradient except in those layers. 
Because of the incompressibility, we have the flow conservation relation
\begin{equation}
f h V_r  =  2\delta_\nu V_r^{-}.\label{eq:v2_3}
\end{equation}
showing that roughly, the top vortex advects from the core up to the top walls, a quarter of the core flow ($\delta_\nu \ll h/4$). It is the same for the bottom roll to the bottom wall.
Concerning the perpendicular flow, they are a narrow recirculation layers where the curvature and viscous forces are balanced: 
\begin{equation}
\frac{{V_\theta^-}^2}{\bar{r}}\simeq \nu \frac{V_r^{-} }{ L_D^2}.\label{eq:v2_2}
\end{equation}
where $L_D= \sqrt{\frac{h\nu }{2 V_\theta^-}}$. 
Note that we have introduced a second viscous width, $\delta_\nu \ne L_D$, because the force balance is in the gradient localization of the viscous layers. Let us emphasize that the boundary layers are viscous layers, not depending on the Hartmann number, since in this regime, the flow is assumed strong and the magnetic field weak. Thus, there is no concentration of current and flow in narrow Hartmann layers, see Fig.~\ref{fig:Innertialrun}b. %
Gathering the three previous equations, we get the scaling law in the inertial regime:
\begin{equation}
V_\theta^{\mbox{\tiny I-R}}  \simeq  \left(\frac{f I_0B_0\Delta r}{2\pi \rho}\right)^{2/3}\left(\frac{4}{\pi \bar{r} h^2 \nu}\right)^{1/3} .\label{eq:v2_final}
\end{equation}
This law is shown for $I_0$ in Figs.~\ref{fig:Scan_Vt_ref_I} and \ref{fig:Scan_Vt_ref_I_lowB},  and for $B_0$ in Fig.~\ref{fig:Scan_Vt_ref_B_lowI}. We observe  good agreement of this relation with the numerical results. 

%%%%%%%%%%%%%%%%%%%%%%%%%%%%%%%%%
%%%%%%%%%%%%%%%%%%%%%%%%%%%%%%%%%
%
%%%%%%%%%%%%%%%%%%%%%%%%%%%%%%%%%%%%%%%%%%%%%
%%%%%%%%%%%%%%%%%%%%%%%%%%%%%%%%%%%%%%%%%%%%%
\subsection{The influence of the geometry}
\begin{figure}
\begin{center}
\includegraphics[width=0.90 \linewidth]{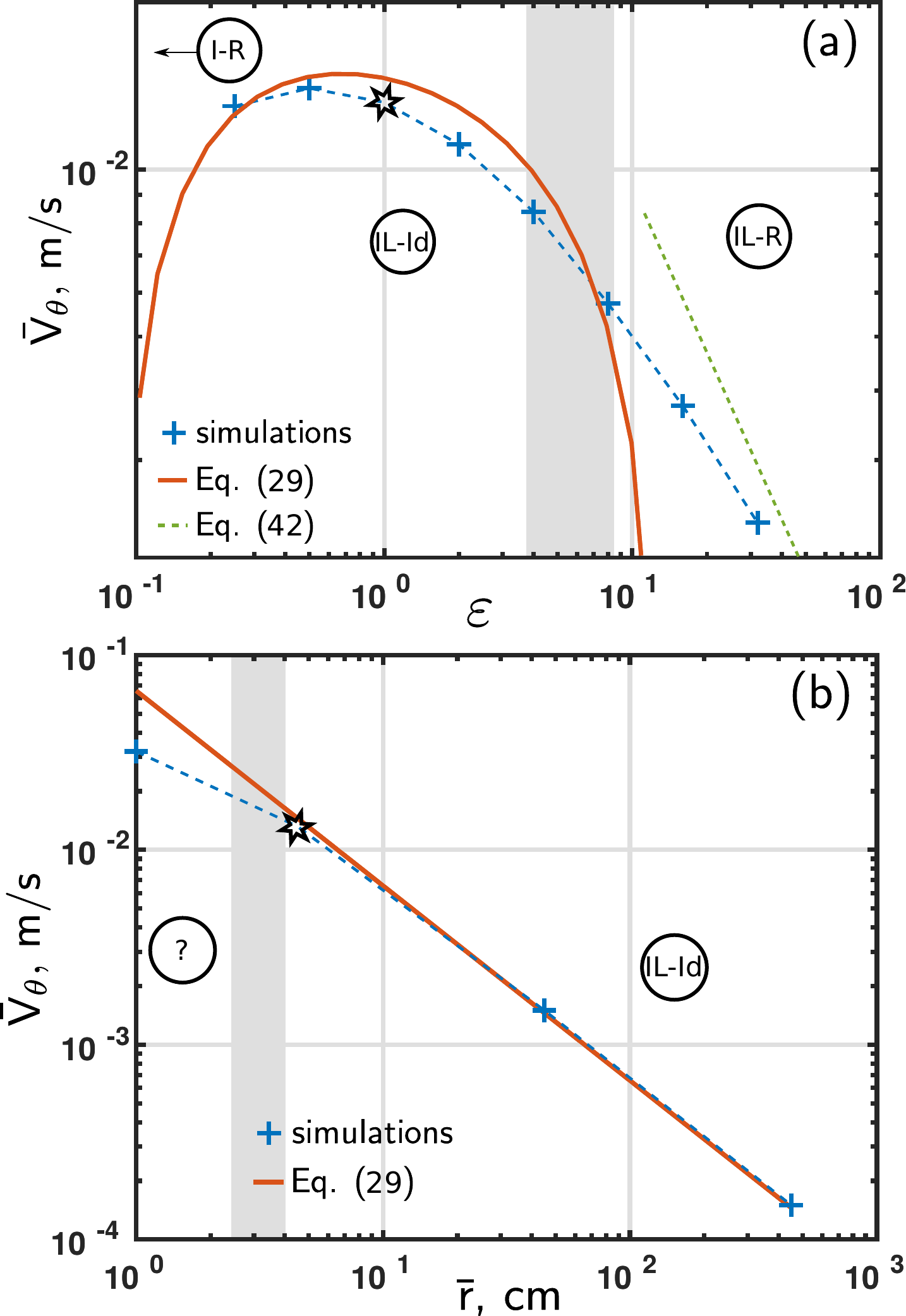}
\caption{  (Color online) Mean toroidal flow for reference case with various aspect ratio values (a) or with various mean radius (b). Blues crosses for simulation. Lines for theories. The scan crosses the identified regimes. The theoretical value Eq.~\eqref{eq:V_0Baylis} in regime \textsc{Il--Id} is the orange line.} \label{fig:Vt_geom}
\end{center}
\end{figure}

%%%%%%%%%%%%%%%%%%%%%%%%%%%%%%%%%%%%%%%%%%%%%
%%%%%%%%%%%%%%%%%%%%%%%%%%%%%%%%%%%%%%%%%%%%%
%
Historically, the experimental ducts \cite{Baylis1964,Moresco2004}  had a square geometry $\epsilon=h/\Delta r=1$ with various mean radius  $\bar{r}\mbox{ [cm]}\in{5.4,6.8}$ and height $h\mbox{ [cm]}\in{0.4,3.1}$. It was also the case in our studies above. However, one may expect that the regimes will depend on the geometric parameters, and potentially some other regimes may appear.  
In fact, we find that the aspect ratio affects the amplitude and profile of $V_\theta$ significantly for the regimes studied. This is shown in Fig.~\ref{fig:Vt_geom}. The points correspond to simulations of the reference case with varying $\varepsilon$ (Fig.~\ref{fig:Vt_geom}a) and $\bar{r}$ (Fig.~\ref{fig:Vt_geom}b). The regimes are again identified by assessing the induction rate value, Eq.~\eqref{eq:tau}, and the ratio $\left<\|(V\cdot\nabla)V\|\right>/ \left<\|\nu \triangle V \|\right >$. The corresponding theoretical laws available are also reported in the graphs. The previous calculation used some simplifications linked to the geometry which do not hold anymore. We have adapted the calculations for the case of the regime \textsc{Il--R} with tall tori. The toroidal velocity is calculated by replacing the assumption $\Delta r << \bar{r}$ by $\partial_z \simeq 0$. Eq.~\eqref{eq:Vt_reg3_1} becomes:
\begin{equation}
\partial_{r}^2 V_\theta(r) +\frac{1}{r}\partial_r V_\theta(r) - \frac{1}{r} V_\theta(r) \simeq - \frac{I_0 B_0}{2h \pi \rho \nu r}.
\end{equation}
The boundary conditions are $V_\theta(r_0) = V_\theta(r_1) = 0$. The value averaged in the complete volume is:
\begin{equation}
V_\theta^{\mbox{\tiny IL-R}} = \frac{I_0 B_0 \bar{r}}{8\pi h \rho \nu}\left(1 - \left(\frac{r_0 r_1}{\Delta r \bar{r}}\ln(r_1/r_0) \right)^2 \right).\label{eq:Vt_long_IL_R}
\end{equation}
The regime \textsc{Il--Id} is not depending on the height of the torus, according to Eq.~\eqref{eq:V_0Baylis}. The parameter $h$ appears explicitly only into a corrective term. However, in (Fig.~\ref{fig:Vt_geom}a), we observe this is not true for large $\epsilon$ or $h$. Indeed, the calculation supposed the presence of viscoresistive layers (Hartmann and Shercliff). It does not remain valid when increasing the aspect ratio: in the \textsc{Il--Id} regime, the Shercliff layer disappears when $h/\sqrt{H\!a} > \Delta r$, thus for torus taller than 25~cm in our reference case.The absence of Shercliff layers allows a radial progression of the current from the inner boundary. Thus, current is not expelled to the top and bottom of the duct, and current field lines are straight and radial. Thus, there is no induction for large enough $\epsilon$ and the system enters in a resistive regime. 
Moreover, top and bottom Hartmann layers do not depend on the height $h$. As explained in section \ref{sec:MeanFlow}, they prevent strong recirculation flows and maintain the inertialess character of the regime when increasing $\epsilon$. We thus have an \textsc{Il--R} regime for tall tori.
Note that, on the contrary, for flat torus or when decreasing $\epsilon$, Hartmann layers overlap and disappear if $1/H\!a > 1$ (h<0.5~mm). We did not investigate numerically those regimes.

Increasing the mean radius of the torus $\bar{r}$ decreases the radial current density, since the total current is inserted across a larger surface. However, the regime remains the \textsc{Il--Id} one's as observed in (Fig.~\ref{fig:Vt_geom}b). 
%

%%%%%%%%%%%%%%%%%%%%%%%%%%%%%%%%%
%%%%%%%%%%%%%%%%%%%%%%%%%%%%%%%%%

%%%%%%%%%%%%%%%%%%%%%%%%%%%%%%%%%%%%%%%%%%%%%

%%%%%%%%%%%%%%%%%%%%%%%%%%%%%%%%%%%%%%%%%%%%%
%%%%%%%%%%%%%%%%%%%%%%%%%%%%%%%%%%%%%%%%%%%%%
\section{Comparison with experiments.}\label{sec:exp}
%%%%%%%%%%%%%%%%%%%%%%%%%%%%%%%%%%%%%%%%%%%%%
%%%%%%%%%%%%%%%%%%%%%%%%%%%%%%%%%%%%%%%%%%%%%
%
\subsection{Tall torus}
In the work of Boisson et al. \cite{Boisson2012, Boisson2017}, the scaling law for $V_\theta$ is different from any of our observations. We recall their result: $V_{B} = 2.1\sqrt{\frac{I_0 B_0 \Delta r}{r_1 r_0 \rho}}-0.015$~m/s. We provide a comparison with their experimental results obtained using Galistan. Simulations are done with their experimental parameters: $r_0 = 1$~cm, $r_1 = 4$~cm, h=12~cm, $\eta=2.89\cdot 10^{-7}\Omega\mbox{m}$, $\rho=6440\ \mbox{kg m}^{-3}$ and $\nu=3.73\cdot 10^{-7}\ \mbox{ms}^{-2}$.  The magnetic field is also varying from 0.01 to 0.1~T and the current from 1~mA to 20~A. 

\begin{figure}
\begin{center}
\includegraphics[width=1.0 \linewidth]{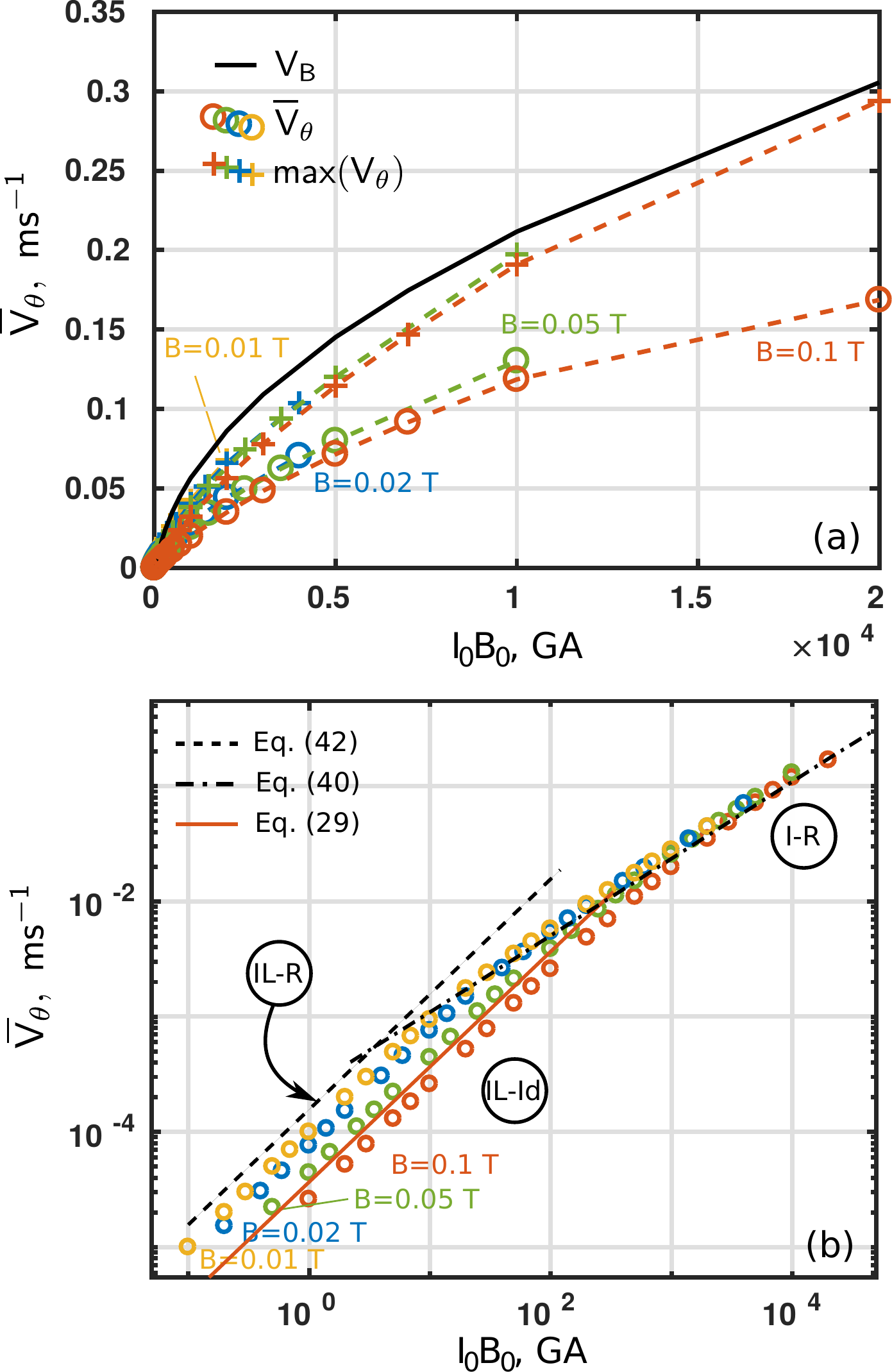}
\caption{ (Color online) Values of toroidal flow versus $I_0B_0$. The black line is the scaling of  Boisson et al. \cite{Boisson2012, Boisson2017}. The crosses are the maximum value of the toroidal flow. The circle are the mean value of it. Colors stand for the magnetic field:yellow 10~mT, blue 20~mT, green 50~mT, 2 orange 0.1~T. Parameters: $r_0 = 1$~cm, $r_1 = 4$~cm, h=12~cm, Galistan, $\eta=2.89\cdot 10^{-7}\ \Omega m$, $\rho=6440\ kg m^{-3}$, $\nu=3.73\cdot 10^{-7}\ ms^{-2}$, $B_0$ from 10~mT to 0.1~T and $I_0$ from 1~mA to 20~A. Panel a) : linear scale, Panel b) : logscale with theoretical value of Eq~\eqref{eq:Vt_long_IL_R} (dashed black line), Eq.~\eqref{eq:V_0Baylis} (orange line for $B_0=1$~T) and Eq.~\eqref{eq:v2_final} (dot-dashed black line).} \label{fig:boisson_Vt}
\end{center}
\end{figure}

The  comparison is shown in Fig.~\ref{fig:boisson_Vt}. The first panel is in linear scale: the scaling $V_B$ from the experiment approximately fitting the maximum speed of our profile. This is consistent with their measurement process: the average is done along a z-line at a radial position where the toroidal flow reaches its maximal value. This average overestimates the value of the volume average (see circles in the figure). The second panel is in log scale and allows to assess the derived scaling laws, if we consider that the measured value of $V_B$ corresponds to the mean value and not the maximal value. Obviously, it is difficult to make precise experimental measurement of a power law over a single decade, which was an experimental limitation in the experiments. To improve our understanding, the scaling $V_{B}$ proposed in reference \onlinecite{Boisson2017} is compared to  our scaling laws, fitting well the numerical results obtained for the regime they studied experimentally in  \cite{Boisson2017}. 
To emphasize this, we have drawn the theoretical functions $\bar{V}_\theta(I_0B_0)$ for the different regimes together with the numerical data. In [\onlinecite{Boisson2012}], they had presented the same graph using experimental data. The \textsc{Il--Id} regime curve is shown for $B=0.01$T (red line) and fits well the data. When increasing the magnetic field, the data shift towards the \textsc{Il--R} regime and the corresponding theoretical \textsc{Il--R} curve (black dashed line). The last regime \textsc{I--R} corresponds to high current and here to high $I_0 B_0$ values. The agreement with the Eq.~(\ref{eq:v2_final}) is correct with a geometrical modification. Indeed, for such tall geometry, the Dean's roll are located on the very top and bottom of the duct. The size of the vortices becomes negligible compare to $h$ and  the radial outward flow extends over almost the full height of the channel, which modifies $f$ to 1 in Eq.~\eqref{eq:v2_3}. 
By mixing their results with the parameter $I_0 B_0$, the authors in [\onlinecite{Boisson2012}] have mixed the different regimes which blurs the transition between the different  power laws.

\subsection{Square torus}\label{sec:exp_square}

%%%%%%%%%%%%%%%%%%%%%%%%%%%%%%%%%
%%%%%%%%%%%%%%%%%%%%%%%%%%%%%%%%%

\begin{figure}
\begin{center}
\includegraphics[width=1.0 \linewidth]{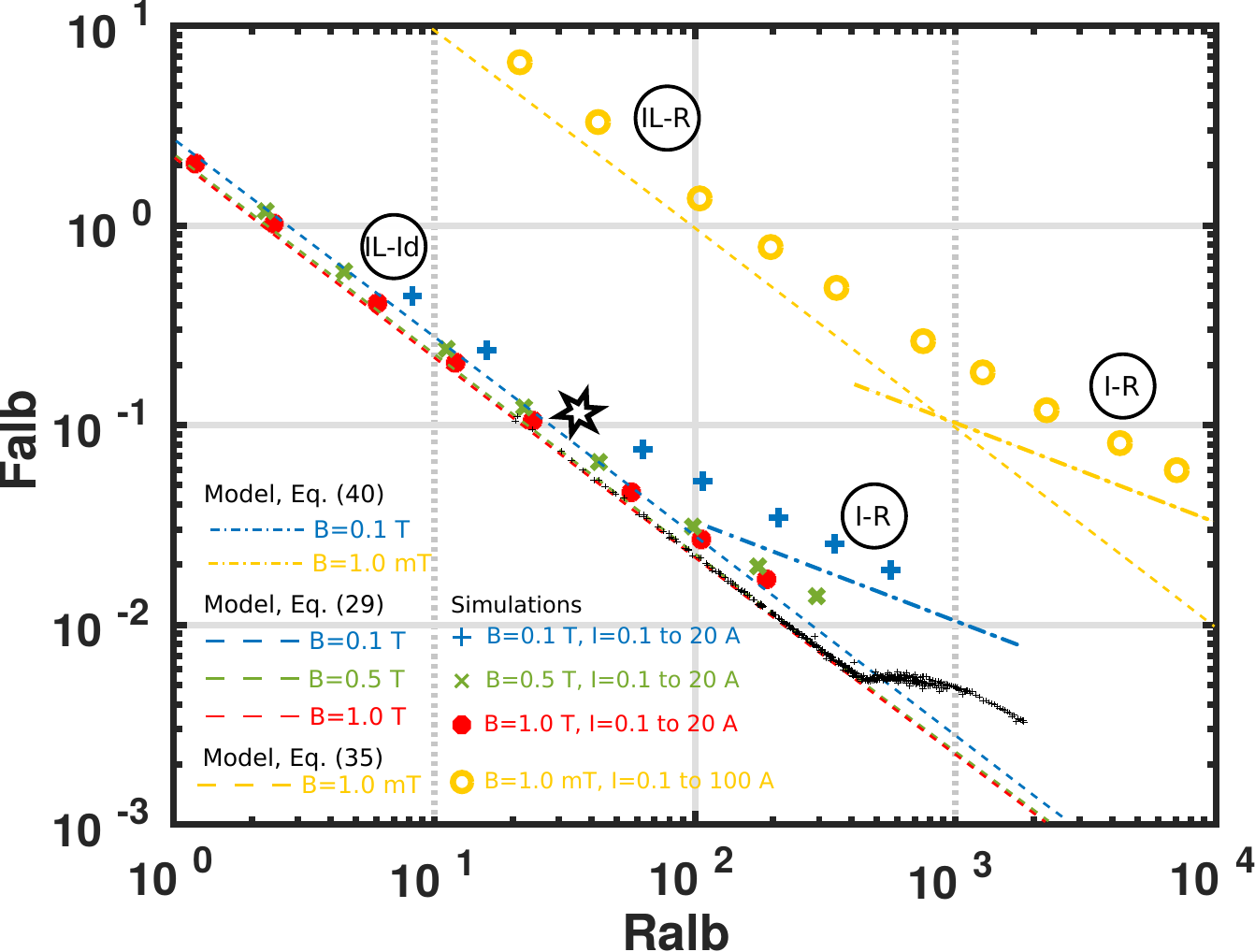}
\caption{ (Color online) The friction factor $F_{Alb}$ in function of $R_{Alb}$. Black thin cross: experimental measurements (1 to 13 T and 10 to 400 A). Clue crosses: simulations with $B_0$ = 0.1~T and $I_0$ = 0.1 to 20 A. Green crosses: simulations with $B_0$ = 0.5~T and $I_0$ = 0.1 to 20 A. Red dot: simulations with $B_0$ = 1~T and $I_0$ = 0.1 to 20 A.  Yellow circle: simulations with $B_0$ = 1~mT and $I_0$ = 0.1 to 100 A. Dashed lines: theoretical values calculated with Eq.~\eqref{eq:V_0Baylis} for blue, green and red and with Eq.~\eqref{eq:Vt_square_IL_R} for yellow. Dot-dashed lines: theoretical values calculated with Eq.~\eqref{eq:v2_final}. Black star: reference case.} \label{fig:alb_exp}
\end{center}
\end{figure}

Moresco and Alboussière \cite{Moresco2004} measured the mean toroidal flow $V_\theta$ obtained for several values of magnetic field and current in the device described in  Fig.~\ref{fig:alb_manip}. Their main result is reproduced in Fig.~\ref{fig:alb_exp}. It represents the friction factor $F_{Alb} = \frac{I_0B_0}{V_\theta^2\rho 2 \pi \bar{r}}$, with $\bar{r} = (r_1+r_0)/2$ the mean radius versus the Hartmann Reynolds $R_{Alb} = R\!e/H\!a=\frac{L V_\theta}{\nu}\frac{\sqrt{\rho\eta\nu}}{LB_0}$. The  friction factor is the ratio between the mean Laplace force and the inertial radial force at the center. The graph, keeping other quantities constant, links the total current $I_0$ to the mean flow $\bar{V}_\theta$.
%
%
%%%%%%%%%%%%%%%%%%%%%%%%%%%%%%%%%
%%%%%%%%%%%%%%%%%%%%%%%%%%%%%%%%%
%
%
Indeed, the experimental data (thin black crosses) are obtained in [\onlinecite{Moresco2004}] with current and  magnetic field ranging respectively from 1 to 400 A and from 1 to 13 T. In this experiment the Shercliff layers are very small compared to the duct dimension ($\delta_{Sh}/(r_1-r_0)$ which varies from  $0.017$ to $0.062$, which explains that the Baylis formulae corrections are negligible. Owing to numerical limitation, we have used weaker magnetic fields. We have compared the full formula Eq.~\eqref{eq:V_0Baylis} (the colored dashed lines on the Fig.~\ref{fig:alb_exp} for each value of the magnetic field) with the value obtained from numerical simulations (colored signs). We also reproduce in the graph the data corresponding to a weak magnetic field of 1~mT and its associated theoretical estimate Eq.~\eqref{eq:Vt_square_IL_R} and Eq.~\eqref{eq:v2_final}. The theories, experiments and numerical simulations match correctly.

\begin{figure}
\begin{center}
\includegraphics[width=0.90 \linewidth]{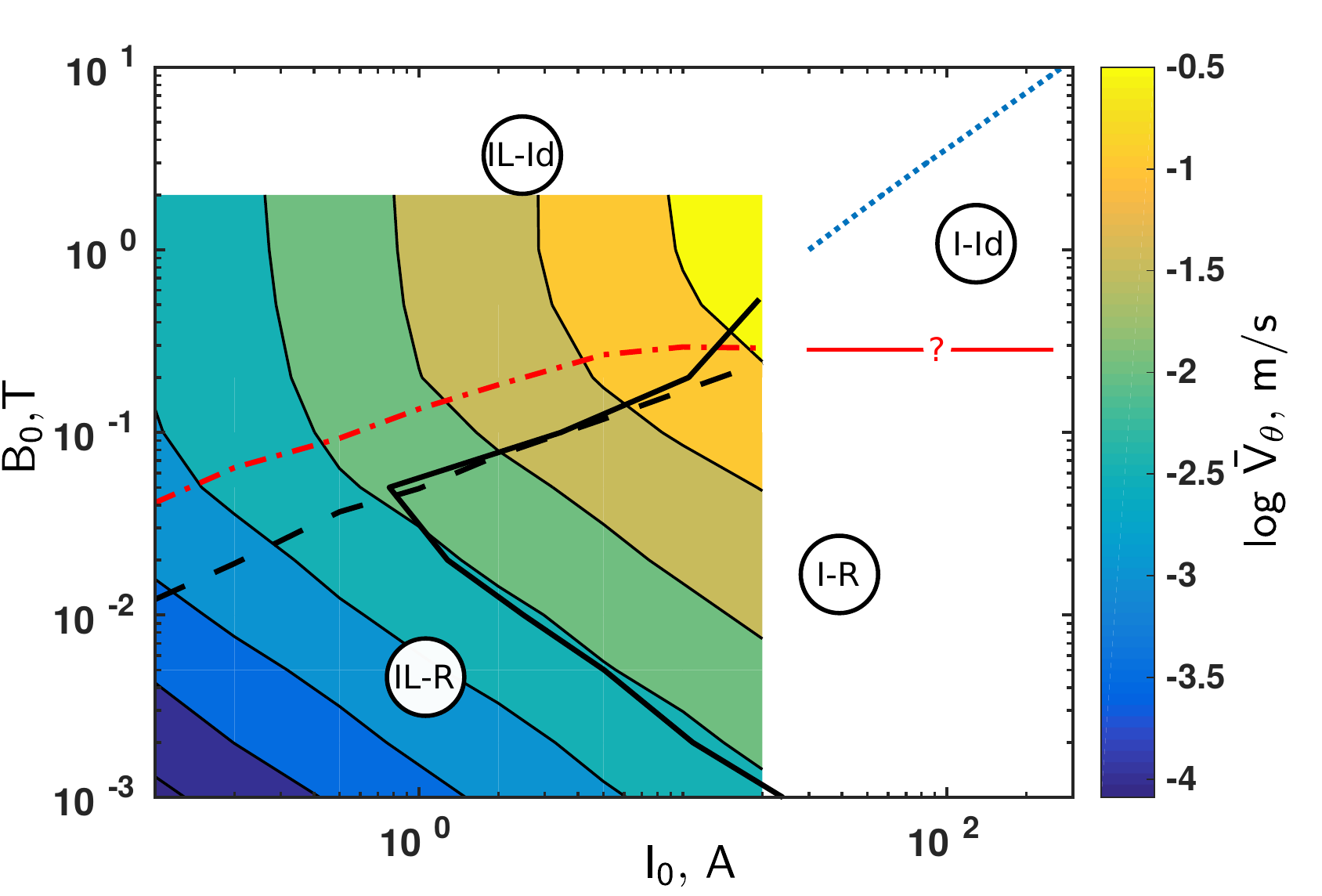}
\caption{ (Color online) The figure \ref{fig:regime} plus extrapolation from experiment (blue dashed line) and a question mark for the limit of the ideal or resistive Ohm's law (red line).}\label{fig:map_IB_2}
\end{center}
\end{figure}

The experimental results \cite{Moresco2004}, see Fig.~\ref{fig:alb_exp}, triggered studies focusing on the problem of the regime change when $R_{alb}\simeq 400$ \cite{Lingwood1999,Krasnov2004,Khalzov2006,Thess2007,Krasnov2010a,Krasnov2010b,Zhao2011}. The global conclusion is that this modification is due to a switching from an essentially 2D-dynamics (r-z), including instabilities and the nature of the turbulence to a 3D dynamical behavior amplified by an intrinsically 3D instability. 
%As obtained by P. Moresco \cite{Moresco2003}, 
For several combinations of values of $B_0$ and $I_0$, it was observed\cite{Moresco2003} that the value at which the flow transitions to another state is always at the same point $ (R_{alb,c}, F_{alb,c})\approx (400,5.10^{-3})$ in the diagram $(R_{alb}, F_{alb})$. %This tends to indicates that it is linked to those adimensional numbers.
From our study it is clear that if $R_{alb} < R_{alb,c}$, the agreement between the experiment and the theory Eq.~\eqref{eq:V_0Baylis} illustrates that the regime is inertialess and ideal (\textsc{Il--Id}).  %
As the transition occurs in the \textsc{Il--Id} regime, we can determine the position of the transition in the  $(I_0,B_0)-$plane. This is  the blue dotted line in Fig.~\ref{fig:map_IB_2}, which results from the condition $(R_{alb}, F_{alb})=(R_{alb,c}, F_{alb,c})$ and the validity of Eq.~\eqref{eq:V_0Baylis} for the  \textsc{Il--Id} regime.

We observe that the transition seems to correspond to the extension of the full black line which divides inertialess and inertial zones. Taking into account also the red line which divides resistive from ideal zones, we can infer that, in the regime they have studied, the transition they observe could correspond to a transition from an inertialess ideal regime to an inertial ideal one, the transition occurring for large enough current. 
Let us emphasize, however, that this claim is not incompatible with the presence of 3D mechanisms and/or turbulence. The extension of the present work to non-axisymmetric flow is in this light an interesting perspective.

%%%%%%%%%%%%%%%%%%%%%%%%%%%%%%%%%%%%%%%%%%%%%
%%%%%%%%%%%%%%%%%%%%%%%%%%%%%%%%%%%%%%%%%%%%%
\section{Conclusion}
\label{sec:concl}
%%%%%%%%%%%%%%%%%%%%%%%%%%%%%%%%%%%%%%%%%%%%%
%%%%%%%%%%%%%%%%%%%%%%%%%%%%%%%%%%%%%%%%%%%%%
%
%%%%%%%%%%%%%%%%%%%%%%%%%%%%%%%%%%%%%%%%%%%%%
%%%%%%%%%%%%%%%%%%%%%%%%%%%%%%%%%%%%%%%%%%%%%

We performed a study of the annular flow driven by an electric field. In the first part of this work, we have shown that even if the asymptotic flow is turbulent, it has a statistical steady state which is comparable with the results obtained from theoretical time independent solutions. Then, we explored the flow properties by varying four physical parameters: the injected current, the magnetic field, the aspect ratio of the duct and its mean radius. The results demonstrate the intricate interplay of those parameters causing different effects which determine the flow regimes. They show that the usual classification of the flow in inertial/inertialess regimes does not encompass all the possible regimes. 
More precisely, we have identified numerically 3 regimes and verified the powerlaws which characterize their statistical steady states. It appears that, associated with the presence or absence of Hartmann and Shercliff layers and the relative importance of nonlinear terms, it is difficult to classify the regimes with a few adimensional numbers. As experiments have various designs, we have also investigated the influence of the geometric parameters and found that they impact on the nature of the regimes. Finally, taking the parameters of two representative experiments,  we have found that the 3 regimes are present. We have specifically compared our numerical and analytical results with their  experimental results. We have recovered the mean toroidal flow values obtained experimentally and highlight the validity of the data with the powerlaws of the 3 regimes. Concerning the experiments of  Moresco and Alboussière, we found that the regime change they identified by plotting the friction factor versus the Hartmann Reynolds corresponds to a cross-over from the inertialess ideal regime to an inertial regime, and is probably not directly linked to the 3D nature of the turbulence.

The value and shape of the toroidal velocity profile are key-parameters to control MRI experiments. They require a decreasing angular momentum profile. However, the path to the MRI instability is still not understood.
First, there is still an unobserved regime in our parameter range, the  inertial ideal regime. Whether or not, it may produce a favorable toroidal flow in MRI context is thus open. 
Second, the MRI is not the only instability that can be triggered with a decreasing angular toroidal momentum profile. Curvature or hydrodynamic instabilities can be triggered in such context (Rayleigh stability criterion). In our simulations, the Rayleigh criterion is never satisfied in the vicinity of the outer boundary. Suslov {\it et al.}\cite{Suslov2017}, by studying an electrolyte flow in a cylinder layer, obtained the same conclusion.  
Thus, experimental conditions which could avoid this curvature instability must be further investigated to sort out a MRI experiment design.
Third, in duct experiments, the poloidal flow was been observed experimentally \cite{Boisson2017}. It seems that this flow correspond to Dean rolls in inertial regimes \cite{Qin2012}. In our study, we observe they  are linked to the angular momentum profiles. They deserve a separated stud.  Indeed, in hydrodynamics, those secondary flows present a high number of shape and regimes \cite{Yamamoto2006} and lead to various inertial regimes and momentum profiles. Thus, one may also expect such a diversity in magnetohydrodynamic inertial regimes.

\end{document}